\documentclass[manuscript]{aastex}

\usepackage{epsfig}                   
\usepackage{multirow}
\usepackage{natbib}
\usepackage{txfonts}                  

\shorttitle{On the need of the Light Elements Primary Process
(LEPP)} \shortauthors{1,2,3,4,5}

\begin{document}

\newcommand{\ctan}{$^{13}$C($\alpha$,n)$^{16}$O }
\newcommand{\nean}{$^{22}$Ne($\alpha$,n)$^{25}$Mg }
\newcommand{\ct}{$^{13}$C }
\newcommand{\nq}{$^{14}$N }
\newcommand{\odo}{$_\odot$}

\title{On the need of the Light Elements Primary Process
(LEPP)}

  \author{S. Cristallo\altaffilmark{1}
    \affil{1 - INAF-Osservatorio Astronomico di Collurania, 64100 Teramo, Italy}\affil{2 - INFN Sezione Napoli, Napoli, Italy}}
  \and
  \author{C. Abia\altaffilmark{1}
    \affil{1 - Departamento de F\'isica Te\'orica y del Cosmos, Universidad de Granada, 18071 Granada, Spain}}
  \and
  \author{O. Straniero\altaffilmark{2}
    \affil{1 - INAF-Osservatorio Astronomico di Collurania, 64100 Teramo, Italy}\affil{2 - INFN Sezione Napoli, Napoli, Italy}}
  \and
  \author{L. Piersanti\altaffilmark{2}
    \affil{1 - INAF-Osservatorio Astronomico di Collurania, 64100 Teramo, Italy}\affil{2 - INFN Sezione Napoli, Napoli, Italy}}


\date{\today}

\begin{abstract}  Extant chemical evolution models underestimate the
Galactic production of Sr, Y and Zr as well as the Solar System
abundances of s-only isotopes with 90$<$A$<$130. To solve this
problem, an additional (unknown) process has been invoked, the
so-called LEPP (Light Element Primary Process). In this paper we
investigate possible alternative solutions. Basing on Full Network
Stellar evolutionary calculations, we investigate the effects on
the Solar System s-only distribution induced by the inclusion of
some commonly ignored physical processes (e.g. rotation) or by the
variation of the treatment of convective overshoot, mass-loss and
the efficiency of nuclear processes. Our main findings are: 1) at
the epoch of the formation of the Solar System, our reference
model produces super-solar abundances for the whole s-only
distribution, even in the range 90$<$A$<$130; 2) within errors,
the s-only distribution relative to $^{150}$Sm is flat; 3) the
s-process contribution of the less massive AGB stars (M$<$1.5
M$_\odot$) as well as of the more massive ones (M$>$4.0 M$_\odot$)
are negligible; 4) the inclusion of rotation implies a downward
shift of the whole distribution with an higher efficiency for the
heavy s-only isotopes, leading to a flatter s-only distribution;
5) different prescriptions on convection or mass-loss produce
nearly
rigid shifts of the whole distribution.\\
In summary, a variation of the standard paradigm of AGB
nucleosynthesis would allow to reconcile models predictions with
Solar System s-only abundances. Nonetheless, the LEPP cannot be
definitely ruled out, because of the uncertainties still affecting
stellar and Galactic chemical evolution models.
\end{abstract}

 \maketitle

\section{Introduction}\label{intro}

Mass-losing Asymptotic Giant Branch (AGB) stars are the main
source of medium- and long-term gas returned to the interstellar
medium (ISM). For this reason, they allow late episodes of stellar
formation, thus prolonging the star-forming lifetime in many
different Galactic environments. In addition, as a result of a
complex combination of internal nucleosynthesis and deep
convective mixing, the wind of AGB stars is heavily enriched in
both light (C, N, F, Na) and heavy elements. About half of the
isotopes from Sr to Pb are produced by AGB stars in their interior
through a slow neutron capture process called s-process (see,
e.g., \citealt{bussorev}). Moreover, the dust forming in their
cool extended circumstellar envelopes efficiently pollutes the
ISM. Therefore, AGB stars play a fundamental role in the chemical
evolution of galaxies.

In this paper we discuss the evolution of the heavy elements
(A$>$90) in the solar neighborhood. Our main goal is to understand
if the current nucleosynthesis models provide a reliable
evaluation of the ISM contamination by AGB stars. The main and the
strong components of the s-process (A$>$90) are produced by
low-mass AGB stars, typically 1.5$\le $M/M$_\odot \le 3.0$.
Lighter s elements (A$<$90) are mainly synthesized by the
s-process in massive stars during core He burning and shell C
burning (the so-called weak component;
\citealt{ka89,beer92,pigna10}). Massive stars are also responsible
for the r process (rapid neutron capture nucleosynthesis; for a
review see \citealt{snegaco}). Most of the isotopes heavier than
iron are produced by both the s and the r process. However there
exist a few isotopes that cannot receive any contribution from the
r process and, for this reason, are called s-only isotopes. An
s-only isotope with atomic number Z is shielded by the r process
due to the existence of a stable isobar with Z-1 or Z-2. For this
reason, the sequence of $\beta$ decays that occurs at the end of
the r process is interrupted before the s-only nucleus is reached.

Galactic Chemical Evolution (hereinafter GCE) models obtained by
combining the s process contribution of AGB stars (main and strong
component) and massive stars (weak s and r process) have been
studied by \cite{trava99,trava01,travaglio04} and \cite{bista}.
\cite{travaglio04} firstly reported a deficit of the predicted
Solar System abundances of Sr, Y and Zr (about -18\%). These three
elements belong to the first s-process peak in the Solar System
composition, which corresponds to nuclei with magic neutron number
N=50. After analyzing the possible uncertainties in their
nucleosynthesis calculations, they concluded that this deficit
would imply the existence of a missing s-process contribution, the
so-called Light Element Primary Process (LEPP). Note that a
different LEPP has also been invoked to explain the abundances of
a large group of light elements with an important contribution
from the r-process. For instance, \citet{montes07} distinguished
between "solar" and "stellar" LEPP, the latter found in metal-poor
halo stars enriched by an r-process. Our findings are limited to
the main s-process from AGB stars and, thus, we only focus on
s-only isotopes in the solar nebula. Therefore, results presented
in this paper do not provide any hint to certify (or exclude) the
existence of a metal-poor primary LEPP, which could have equally
well
its roots in a sort of weak r-process.\\
The need of an unknown pure s-process contribution has been also
claimed by \cite{bista} basing on the analysis of the s-only
isotopes (see also \citealt{kappeler2011}). Indeed, in their
chemical evolution models all the s-only isotopes with
90$<$A$<$130 are systematically underestimated. As a matter of
fact, the AGB yields used by \cite{travaglio04} and \cite{bista}
are based on post-process calculations \citep{ga98} in which the
main neutron source (the \ct pocket) is artificially introduced.
The s process in low mass AGB stars is mainly due to the neutrons
released by the \ctan reaction in a thin \ct pocket that forms
after each third dredge up (TDU) episode
\citep{stra95,ga98,stra06}. At present, a reliable evaluation of
the extension in mass and of the \ct profile within the pocket is
probably the most challenging task for AGB stellar modelers
\citep{he97,deto03,cri09,cri11,liu13}. In the GCE models by
\cite{bista}, the extension of the $^{13}$C-pocket as well as the
mass fractions of $^{13}$C and $^{14}$N (the main neutron source
and the main neutron poison, respectively) are freely varied in
order to reproduce the 100\% of solar $^{150}$Sm with an s-only
distribution as flat as possible. These authors, however, did not
explore the physical motivation at the base of those variations.
More recently, \cite{trippa} argued that, in stars with M$<$1.5
M$_\odot$, magnetic fields are able to shape larger \ct pockets
than those characterizing more massive AGBs (see also
\citealt{maiorca}) and suggest that this occurrence might have
important consequences on the Solar System s-only distribution.
However, their conclusions have to be verified with the support of
a GCE model as well as that of evolutionary models
that include the feedback of magnetic fields. \\
The analysis of the LEPP problem presented in this work is based
on a different approach. We verify if our FUll Network Stellar
(FUNS, see \citealt{stra06} and references therein) yields,
incorporated into a chemical evolution model for the solar
neighborhood, can provide a reasonable fit to the Solar System
s-only distribution. The adoption of a large nuclear network
directly coupled to the physical evolution of the stars as well as
our handling of the convective/radiative interface at the base of
the convective envelope (i.e. where the \ct forms) do not allow us
to force our calculation to fit the absolute value of the Solar
System s-only distribution. Notwithstanding, we can evaluated the
effects on the AGB nucleosynthesis of different prescriptions for
convective overshoot during the TDU, rotation-induced mixing,
pre-AGB and AGB mass-loss rates and nuclear reactions
efficiencies.

The paper is structured as follow. We firstly describe our
Galactic Chemical Evolution model and the stellar evolutionary
code used to determine the proto-solar distribution for s-only
isotopes. (Section \ref{gce} and Section \ref{modagb},
respectively). Then, in Section \ref{refe} we present our
reference case, while in Sections \ref{agb} and \ref{chiesto} we
describe how AGB models uncertainties and GCE uncertainties affect
our results, respectively. Our conclusions follow in Section
\ref{conclu}.

\section{The Galactic Chemical Evolution Model}\label{gce}
We use a simplified GCE model for the solar neighborhood, defined
as a cylinder  of $\sim 1$ kpc radius at a distance  of $\sim$8
kpc from  the Galactic center, adopting the standard formalism
\citep[e.g.][]{Pagel2009}. Our GCE code is an update of that used
to follow the evolution of light elements in previous studies
\citep{abia91,Abia1995}. The classical set of equations are solved
numerically  without the instantaneous recycling approximation
(i.e. stellar lifetimes are taken into account) and assuming that
at the star's death its ejecta are thoroughly mixed
instantaneously in the local ISM, which is then characterized by a
unique composition at a given time. Thus, our predictions
represent average values in time as this simplified approach
cannot account for the scatter in any Galactic observable. Our
main goal is to reproduce the absolute and relative isotopic
abundances distribution of s-only nuclei at the Solar System
formation (occurred 4.56 Gyr ago). These nuclei are $^{70}$Ge,
$^{76}$Se, $^{80,82}$Kr, $^{86,87}$Sr, $^{96}$Mo, $^{100}$Ru,
$^{104}$Pd, $^{110}$Cd, $^{116}$Sn, $^{122,123,124}$Te,
$^{128,130}$Xe, $^{134,136}$Ba, $^{142}$Nd, $^{148,150}$Sm,
$^{154}$Gd, $^{160}$Dy, $^{170}$Yb, $^{176}$Lu, $^{176}$Hf,
$^{186}$Os, $^{192}$Pt, $^{198}$Hg and $^{204}$Pb. We concentrate
on those isotopes because they are produced only via the s-process
and because an AGB origin is certain for those with atomic mass
$A\ge 96$. We use the absolute isotopic abundances of the
proto-solar nebula to normalize the output of our GCE model, which
is stopped at that epoch. Note that those abundances differ from
the current ones observed in the solar photosphere due to the
impact of chemical settling. We obtain our proto-solar
distribution by adopting the elemental abundances of \cite{lo09}
and by computing a Standard Solar Model according to the procedure
describe in \cite{Piersanti2007}.

The basic ingredients of the GCE model are
described in the following. For the stellar yields see \S \ref{modagb}.\\

We adopt a standard Salpeter Initial Mass Function (IMF),
$\Phi(M)\propto M^{-(1+X)}$ with $X=1.35$ in the mass range
0.1-100  M$_\odot$. For the Star Formation Rate (SFR) we have
adopted the standard Schmidt type law $\Psi(t)= \alpha
\sigma_{gas}^{k}(t)$, where $\sigma_{gas}$ is the surface gas
density, $\alpha=0.32$ Gyr$^{-1}$ and $k=1$. \\
We assume that the disk has been build up starting from an initial
surface gas density ($\sigma_o$)  and by slow accretion of gas
with primordial  composition. Hence the initial abundances of all
the studied isotopes is set to zero. We adopt an exponentially
decreasing  gas accretion law $f(t)\propto e^{-t/\tau}$. This
infall prescription, combined to the adopted SFR law, leads to a
decreasing star formation history in the solar neighborhood. We
set $\tau=7.5$ Gyr since it has been shown that such a long
timescale provides a satisfactory fit to the observed stellar
Metallicity Distribution (hereinafter MD) in the solar
neighborhood \citep[e.g.][]{Boissier1999}. We have normalized the
infall rate $f(t)$ by imposing that the current observed total
surface density is $\sim 50$ M$_\odot$ pc$^{-2}$ \citep[see][and
references therein for a detailed discussion]{Goswami2000}.

The main observables in the solar neighborhood which must be
fitted are: \begin{itemize} \item{the current surface density  of
gas ($13\pm 3$ M$_\odot$pc$^{-2}$; gas fraction 0.15-0.25\%),
stellar    surface density    ($35\pm 5$ M$_\odot$ pc$^{-2}$),
total mass  ($\sim 50$ M$_\odot$ pc$^{-2}$) as well as the current
star formation rate (2-5 M$_\odot$Gyr$^{-1}$pc$^{-2}$)
\citep{Boissier1999,Goswami2000}};\item{the observed
age-metallicity relation \citep[e.g.][]{Casagrande2011};}\item{the
Type II and Ia supernova rates in the Galaxy \citep{Li2011}, and
the observed [O/Fe] vs. [Fe/H] relationship in thin and thick disk
stars \citep[e.g.][]{rami2013,Nissen2014};}\item{the observed MD
of long-lived G-type stars \citep{Casagrande2011, Adibekyan2012,
Bensby2014};}\item{the absolute and relative s-only isotopic
abundances distribution at the formation of the Solar System
\citep{lo09}.}
\end{itemize} Recent studies \citep{Roskar2008,
Schonrich2009, Kubryk2013} have shown that the existence of gas
and star migration across the disk of the Milky Way  can
significantly alter  the local observed age-metallicity relation
and the stellar MD. One of the main conclusions of those studies
is that these observational constraints can be properly
interpreted only if migration of stars and gas is included in GCE
models. For instance, recent GCE models that include migration
show that the average age-metallicity relation for stars locally
born is generally flatter than the one calculated classically
(i.e. without migration, as it is typically done in one zone, 1D
GCE models). In particular, it implies that the Sun was not
probably born locally, but it migrated from inner (more
metal-rich) Galactic regions  up to its current position ($r\sim
8$ kpc). Stellar migration also introduces a dispersion in the
observed abundance ratios as a function of the metallicity (i.e.
[X/Fe] vs. [Fe/H]). Although this dispersion seems to  be
generally small ($\sigma < 0.15$ dex), it might be larger for
elements produced in low-mass, long-lived stars, like Fe or
s-elements. Furthermore, the average gas metallicity in the ISM
might differ from that of the local stellar population. The impact
of the gas and star migration in the observed s-element
distribution in the Solar System is beyond the scope of this
study. We refer to specific studies \citep{Kubryk2013} for a
detailed discussion  on the effects of migration on the chemical
evolution of the Galaxy.\\
Note that we cannot {\it a priori} exclude that the
simplifications inherent our GCE code mask some chemical features
or introduce some biases in following the chemical evolution of
s-only isotopes. Indeed, we are aware that more sophisticated
models than the one zone GCE approximation adopted in the present
work can be constructed for the solar neighborhood. We refer, for
instance, to works including the evolution of the halo and of the
thick disk \citep{Goswami2000,Kobayashi2000,Micali2013}. Different
prescriptions for the SFR and infall/outflow of gas are typically
adopted for the evolution of these two Galactic structures, which
are mainly constrained by their observed MD function. Nonetheless,
we have checked that our results for the s-only isotopes abundance
distribution are not affected when adding, for instance, the halo
evolution (according to the \citealt{Goswami2000} prescriptions),
provided that the initial metallicity for the disk evolution does
not significantly exceed [Fe/H]$\sim -1.0$. Typically, this is the
maximum value of the metallicity for the halo reached in most GCE
models after t$\sim 1$ Gyr.\\

\section{Stellar Models}\label{modagb}

Stellar lifetimes, remnants masses and yields for low  and
intermediate mass stars (1.0$\le$ M/M\odo ~$\le$ 6.0) are derived
from theoretical evolutionary models computed with the FUNS
evolutionary code \citep{stra06}\footnote{The FUNS code has been
derived from the FRANEC code \citep{chi89,chi98}.}. The stellar
yields have been obtained by evolving models with different masses
and initial chemical composition from the pre-Main Sequence up to
the AGB tip. In our models, the adopted AGB mass-loss rate has
been calibrated on the period-luminosity and period-mass loss
relations observed in Long Period Variable Stars \citep[see][and
references therein]{stra06}. The atomic and molecular opacities in
the cool envelope of AGBs account for the variation of the
chemical composition as due to the occurrence of recurrent TDU
episodes \citep{cri07}. During TDU episodes, the instability
occurring at the inner border of the convective envelope is
handled by adopting an exponential decay of the convective
velocities. This makes the TDU deeper; moreover, as a by-product,
we obtain the self-consistent formation of the \ct pocket after
each thermal pulse (TP) followed by TDU \citep{cri09,stra14}. The
extension of the \ct pocket varies from TP to TP following the
shrinking of the region between the He and the H-shells (the so
called He-intershell).
\begin{figure*}[tpb]
\centering
\includegraphics[width=\textwidth]{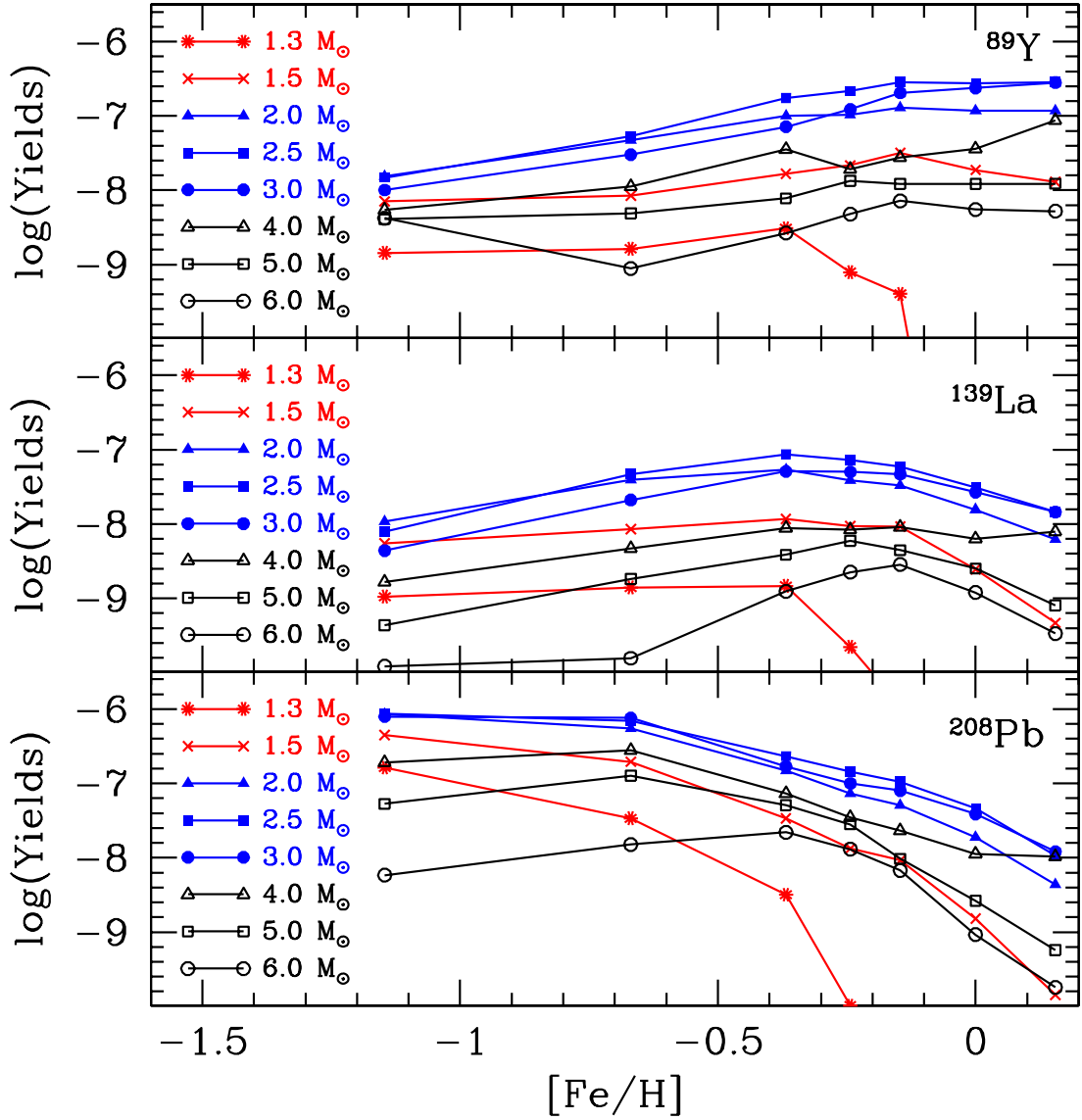}
\caption{Selection of FRUITY yields for key s-process elements.
Upper panel: $^{89}$Y (representative of the first s-process
peak); intermediate panel: $^{139}$La (representative of the
second s-process peak); lower panel: $^{208}$Pb (representative of
the third s-process peak). See the on-line edition for a colored
version of this figure.} \label{fig0}
\end{figure*}
The nuclear network used to follow the physical and chemical
evolution of our models has been presented in \cite{cri11}: it
includes about 500 isotopes (from hydrogen to bismuth) linked by
more than 1000 reactions. Such a network is directly included in
the FUNS code, thus avoiding
the use of post-process techniques.\\
The main neutron source in AGB stars is represented by the \ctan
reaction, burning in radiative conditions at T$\sim 10^8$ K during
the inter-pulse phases. An additional contribution comes from the
activation of the \nean reaction at the base of the convective
shells generated by TPs when the temperature exceeds $3\times
10^8$ K. While the former reaction dominates the s-process
nucleosynthesis in low mass AGB stars, the latter becomes fully
efficient in stars with M$\ge$ 3 M\odo\footnote{This mass limit
depends on the metallicity. As a general rule, the minimum mass
decreases with the metallicity.}.\\
The models we use to calculate the AGB yields have different
masses (1.0$\le$ M/M\odo ~$\le$ 6.0) and metallicities
(-2.15$<$[Fe/H]$<$+0.15; \citealt{cri09,cri11}; Cristallo et al.
\emph{in preparation}). The corresponding yields are available
on-line on our web repository
FRUITY\footnote{http://fruity.oa-teramo.inaf.it}, which represents
our reference set. \\
In Figure \ref{fig0} we report a selection of FRUITY net
yields\footnote{A net yield is defined as \begin{equation}
\int_0^{\tau(M_i)}[X(k)-X^0(k)]\frac{dM}{dt}dt
\end{equation}
where $dM/dt$ is the mass loss rate, while $X(k)$ and $X^0(k)$
stand for the current and the initial mass fraction of the
$k$-isotope, respectively. } for some key s-process elements
($^{89}$Y as representative of the first s-process peak,
$^{139}$La as representative of the second s-process peak and
$^{208}$Pb as representative of the third s-process peak). As
already remarked in \cite{cri11}, the largest yields are produced
in the (1.5-3.0) M$_\odot$ mass range. Figure \ref{fig0} shows
that low mass models (M$<$1.5 M$_\odot$) marginally contribute to
the global s-process production, since the TDU practically ceases
to occur when the initial stellar mass drops below 1.2 M$_\odot$
(see also \S \ref{reimers}). Similarly, s-process yields from more
massive AGBs (M$>$4.0 M$_\odot$) are low, even if these stars may
significantly contribute to the nucleosynthesis of some neutron
rich isotope (as, for example, $^{87}$Rb and $^{96}$Zr) due to the
activation of the \nean reaction. As expected, for stars with
masses between 1.5 and 3.0 M\odo, the relative distribution of the
three s-process peaks weakly depends on the mass, while it has a
different behavior depending on the initial iron content. At large
metallicities ([Fe/H]$>$-0.3), the s-process mainly populates the
first peak (Sr-Y-Zr region). At intermediate metallicities, the
second s-process peak (Ba-La-Ce-Nd region) presents its maximum.
At low metallicities ([Fe/H]$\lesssim$-0.7) lead production
dominates.

In our GCE we adopt a simplified prescription by assuming that all
stars with mass M $>8$ M$_\odot$~explode as core collapse
supernovae leaving behind a compact remnant such a neutron star of
mass $\lesssim$ 1.4 M$_\odot$, or a black hole in the case of most
massive stars (M $>40$ M$_\odot$). In our calculations, we do not
include the contribution of massive stars to the s-process
inventory. Those stars largely contribute to the production of
s-only isotopes with $A\le 87$ \citep[see][and references
therein]{pigna10}. Therefore, for those isotopes our prediction
have to be considered as lower limits. Oxygen and iron yields from
massive stars are instead needed in order to reproduce the average
[O/Fe] vs. [Fe/H] relationship observed in unevolved stars
\citep[e.g.][]{rami2013,Nissen2014}. For that purpose, we use the
yields published by \cite{Chieffi2004}. As far as it concerns the
core-collapse supernovae contribution to the iron enrichment, we
assume that on average each supernova ejects 0.1 M$_\odot$ of
$^{56}$Fe. On the other hand, we adopt the type Ia supernovae
explosion rate according to \cite{greggio1983} in the framework of
the Single Degenerate scenario for their progenitors. This
corresponds to assume that a fraction of $2.5\%$ of all binary
system ever formed in the adequate mass range will provide an
explosive outcome. This fraction value is set by fitting the
observed current Galactic SN Ia rate and the [O/Fe] vs. [Fe/H]
relationship. We also assume that, on average, for each SNIa event
an amount of  $\sim 0.7$ M$_\odot$ of $^{56}$Fe is ejected
\citep[e.g.][]{Bravo2012}.

\begin{figure*}[tpb]
\centering
\includegraphics[width=\textwidth]{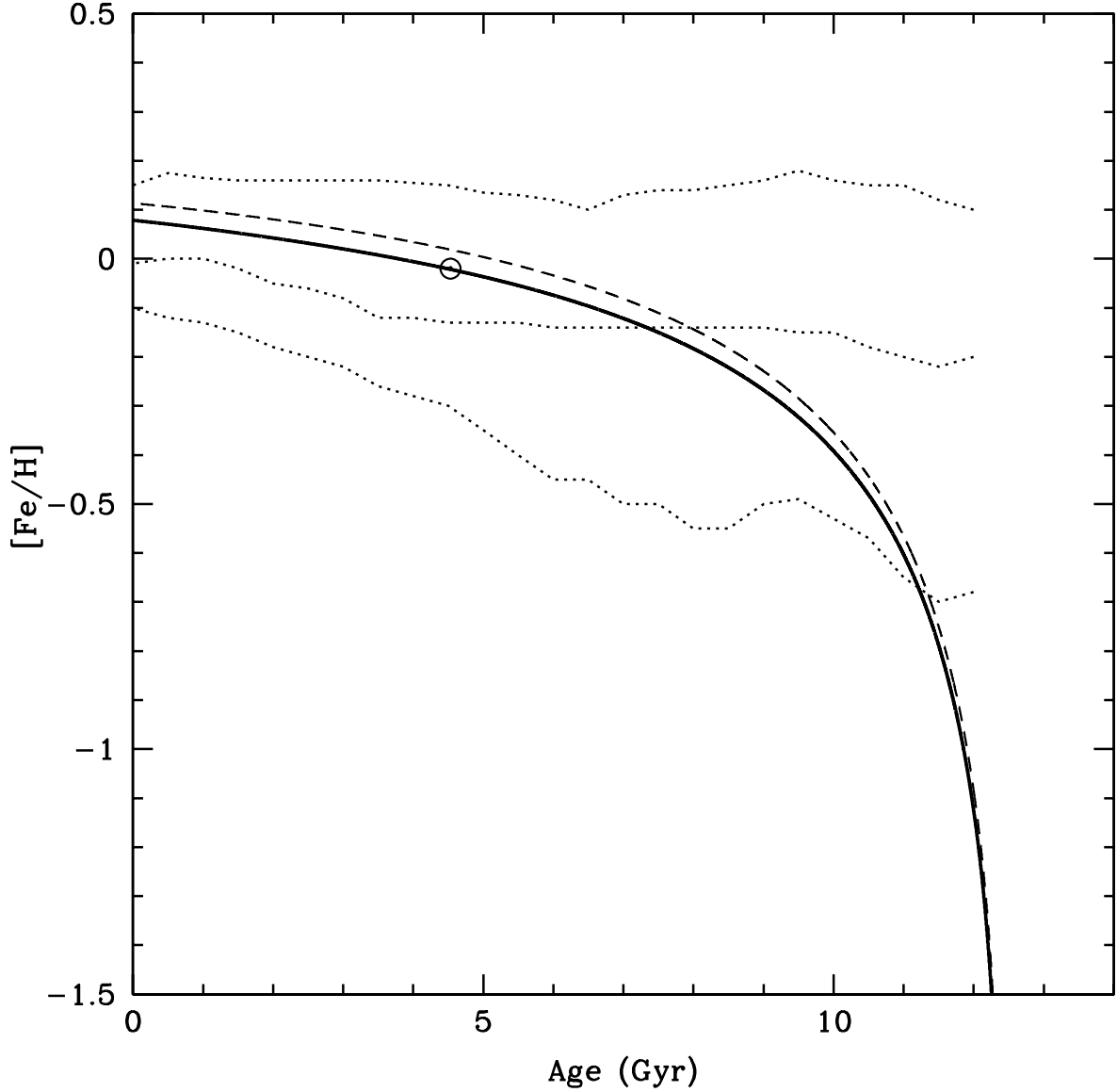}
\caption{Age metallicity relation (solid thick curve) compared to
the average and $\pm1\sigma$ limits (dotted curves) of the
observations in the solar neighborhood by \cite{Casagrande2011}
and \cite{Bensby2014}.\ The dashed curve refers to a GCE model
computed with an increased SFR (+10\% with respect to the solid
thick curve). See Section \ref{chiesto} for details.  }
\label{figc1}
\end{figure*}

\section{Reference case}\label{refe}

Our {\it Reference} case has been computed using the GCE model
described in \S \ref{gce} by adopting parameters values reported
there. The model accounts for all the constraints mentioned above
within the observational uncertainties. It is very well known that
other reasonable choices of the GCE model parameters (SFR law,
IMF, etc.) might give similar results still in good agreement with
the observational constraints. Due to its relevance for our
discussion, we show in Fig. \ref{figc1} the age-metallicity
relation obtained in the {\it Reference} case (thick continuous
line).  The dotted curves represent the average and $\pm 1\sigma$
limits of the observations of \cite{Casagrande2011} and
\cite{Bensby2014}. Our model predicts a rapid increase of  the ISM
metallicity  with time, reaching [Fe/H]$\approx  0.0$ at the epoch
of Solar System formation and a continuous increase of [Fe/H]
until now.

\begin{figure*}[tpb]
\centering
\includegraphics[width=\textwidth]{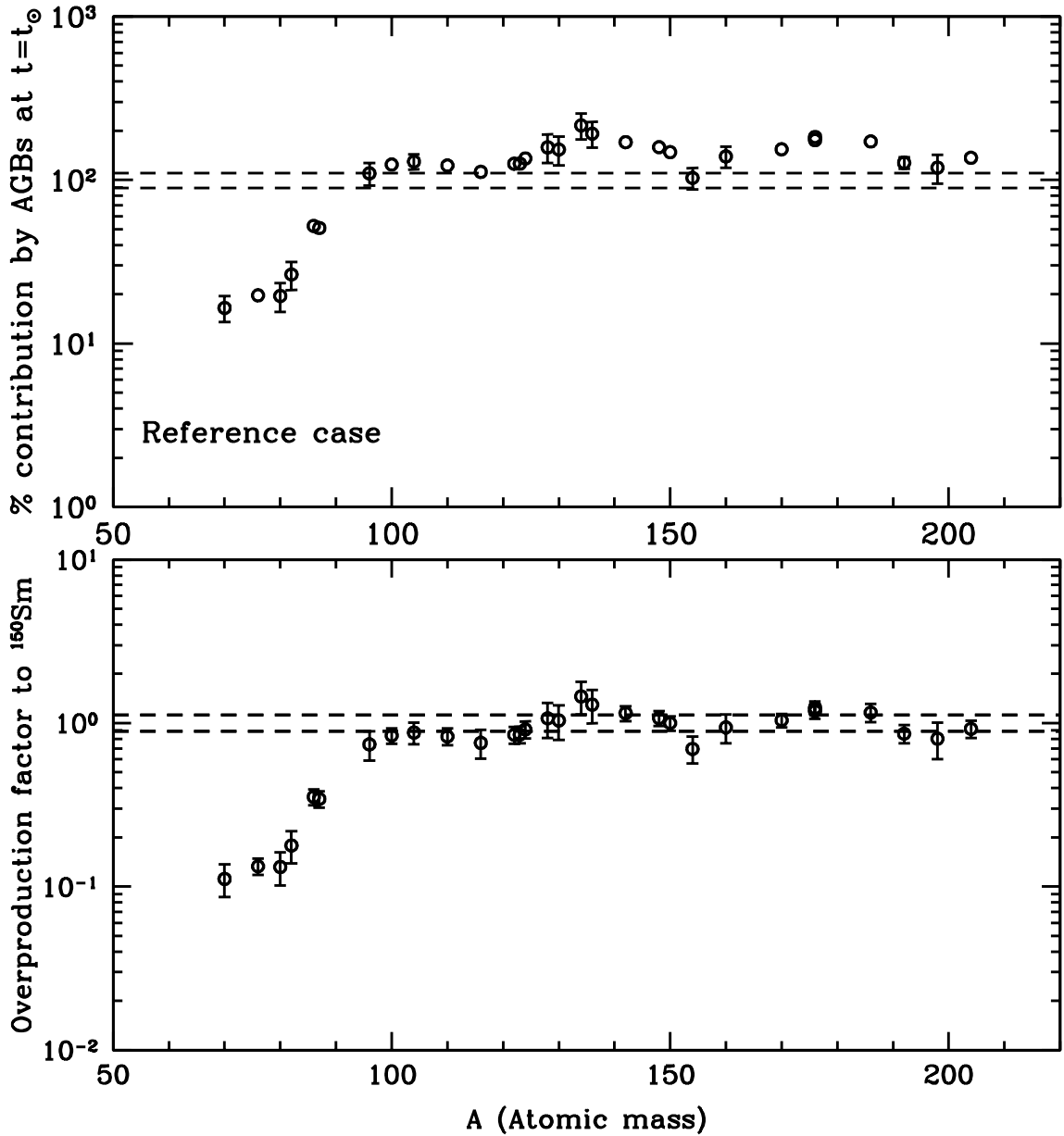}
\caption{Reference GCE calculation case. Upper panel: percentage
s-only isotopic abundances at the epoch of the Solar System
formation. Solar abundance errors are taken from \cite{lo09} and
reported in Table \ref{tab2} (error bars smaller than symbols are
not reported). Lower panel: overproduction factors normalized to
the $^{150}$Sm one. Dashed horizontal lines identify a $\pm$10\%
tolerance region in both panels. See text for details.}
\label{fig1}
\end{figure*}

\begin{figure*}[tpb]
\centering
\includegraphics[width=\textwidth]{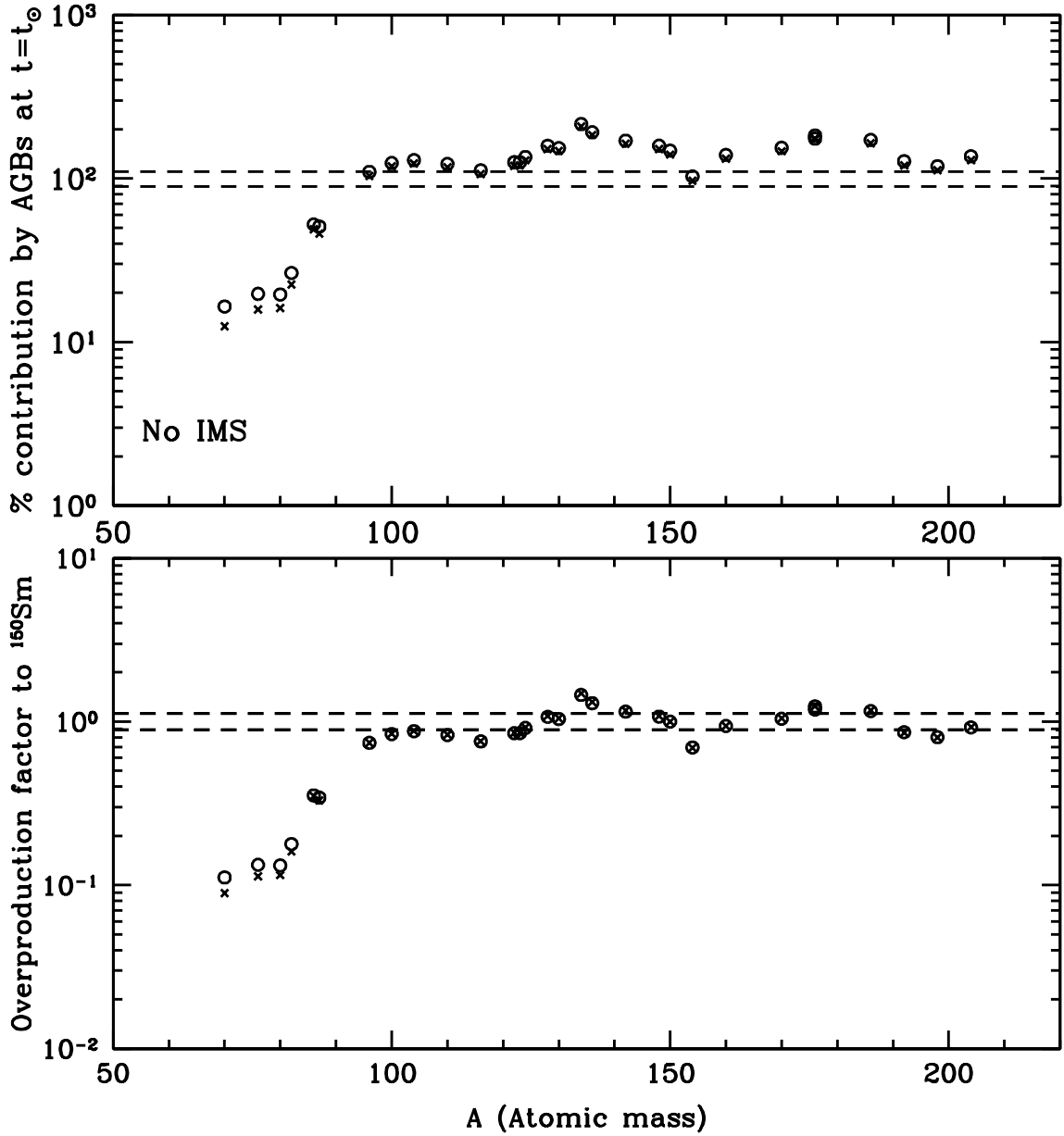}
\caption{As in Figure \ref{fig1}, but including a GCE calculation
without the contribution from IMS AGBs (crosses). The {\it
Reference} case (open dots) is shown by comparison. We have
omitted error bars for clarity.} \label{fig6}
\end{figure*}

In Fig. \ref{fig1}  we report the results of our {\it Reference}
GCE calculation case, as obtained by using stellar yields included
in the FRUITY database. Corresponding data are reported in Table
\ref{tab2}. In the upper panel we report absolute percentage
s-only isotopic abundances obtained from our GCE model at the
epoch of the Solar System formation. In this plot we did not add
any contribution from the weak-s process. Thus, 100\% means that
an isotope is entirely synthesized by AGB stars. For each isotope,
we also plot the corresponding solar abundance uncertainty
reported in \cite{lo09}. Dashed horizontal lines identify a
$\pm$10\% tolerance region representing the current uncertainties
in the estimated chemical abundances as due to nuclear cross
sections \citep{kappeler2011}. In the lower panel of Fig.
\ref{fig1} we report the overproduction factors normalized with
respect to $^{150}$Sm. In this case, unity means that an s-only
isotope is over-produced (or under-produced) as $^{150}$Sm with
respect to the corresponding solar abundance. The latter isotope
has been chosen as reference since the entire s-process flux
passes through it, making this isotope virtually unbranched
\citep{arla99}. Also in this case,
we highlight a $\pm$10\% tolerance region.\\
An inspection to Fig. \ref{fig1} (upper panel) reveals an overall
overproduction of s-only isotopes with A$\ge$96 ($\sim$ 145\%),
more evident in the region 128$\le$A$\le$204. Thus, on a relative
scale, lighter s-only isotopes are underproduced with respect to
the heaviest ones (see lower panel of Figure \ref{fig1}). However,
our relative distribution can be considered flat if current
uncertainties (observational and nuclear) are taken into account.
Since we do not assume any \emph{ad hoc} re-scaling of the \ct
pocket, at odds with \cite{travaglio04} and \cite{bista} who
claimed a missing contribution to light s-only isotopes, we obtain
super-solar percentages for all s-only isotopes with a sure AGB
origin (A$\ge$96). Thus, in the following we investigate if there
is the possibility to decrease the overall Galactic s-only
production and if a larger depletion efficiency can be found for
the heavier s-only isotopes (128$\le$A$\le$204). This exploration
is carried out in the next Section by studying current
uncertainties
affecting stellar models.\\
As starting point of our analysis, however, we want to verify if
GCE models confirm that the bulk of the s-process comes from AGB
stars with masses 1.5$\le$M/M$_\odot \le$3.0 (see previous
Section). The contribution from AGBs with M$<$1.5 M$_\odot$ will
be analyzed in \S \ref{reimers}. In order to quantify the
contribution to the Solar System s-only distribution from
Intermediate Mass Stars AGBs (i.e. stars with initial masses M$>4$
M$_\odot$, hereinafter IMS-AGBs; see also \citealt{kl14}), we run
a GCE model by setting to zero the yields of those objects
(hereinafter No IMS case). Results are shown in Figure \ref{fig6}
and reported in Table \ref{tab2}. On average, the IMS contribution
to the Solar System s-only distribution is marginal (on average
6\%). Thus, even if our IMS-AGBs present tiny $^{13}$C-pocket
after TDUs \citep{stra14}, their contribution, once weighted on
the IMF, is small. For the lightest s-only isotopes (form
$^{70}$Ge to $^{87}$Sr), the relative IMSs contribution is larger,
due to the more efficient activation of the
$^{22}$Ne($\alpha$,n)$^{25}$Mg source. Note, however, that those
isotopes are mainly synthesized by the weak-s component
\citep{kapp94,pigna10}. Thus, we basically confirm the finding of
\cite{bista} that intermediate mass AGBs marginally contribute to
the Galactic chemical evolution of s-only isotopes.

\section{Stellar Models Uncertainties}\label{agb}

Despite great strides made by stellar modelers in the last two
decades, our understanding of the AGB phase is still hampered by
large uncertainties. A few physical details can be constrained by
theory and, therefore, the adoption of phenomenological models is
often the only way to describe a specific physical process. Thus,
it is not surprising that a large part of the extant models still
include a set of parameters sometimes rather free, sometimes
(partially) constrained by observations. Despite all these
limitations, we try to evaluate the effects that current stellar
modelling uncertainties have on s-process yields. In the present
work we focus on some key physical processes (such as rotation,
convection and mass-loss) and on the efficiency of some nuclear
processes.

Due to the large number of models included in the FRUITY database,
we compute a reduced number of (M,Z) combinations by analyzing
once a time each of the above mentioned physical processes and we
derive corrective factors to be applied to models with similar
masses (M) and metallicities (Z). Such a procedure does not
introduce biases in our conclusions because we focus our attention
on those (M,Z) combinations where, according to our previous
experience, major effects are expected.

\subsection{Rotation}\label{rotazione}

\begin{figure*}[tpb]
\centering
\includegraphics[width=\textwidth]{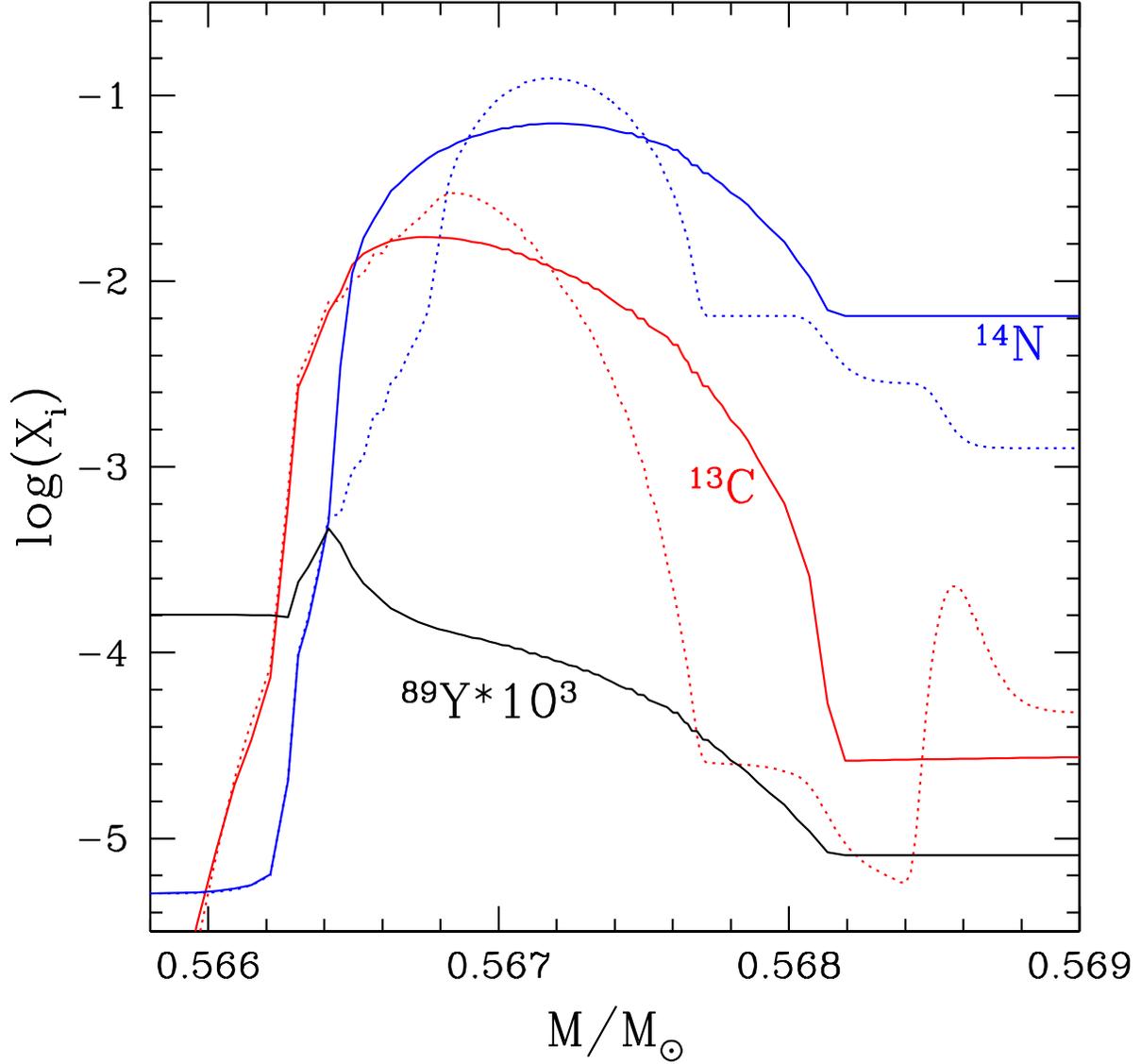}
\caption{$^{13}$C and $^{14}$N profiles after the 4$^{th}$ TDU in
a 2 M\odo~rotating model with Z=10$^{-2}$ ([Fe/H]=-0.15) and
v$^{rot}_{ZAMS}$=30 km/s (solid curves). The $^{89}$Y profile
(multiplied by a factor 1000) is also displayed. We show $^{13}$C
and $^{14}$N profiles at the end of the formation of the \ct
pocket (dotted curves) and when neutrons start being released
(solid curves). See the on-line edition for a colored version of
this figure.} \label{fig7}
\end{figure*}

FRUITY AGB stellar models are representative of the intrinsic
carbon stars observed in the disk and in the halo of the Milky
Way. However, a comparison between our theoretical curves and
spectroscopic data shows that, at fixed metallicity, our models do
not cover the observed spread in the s-process indexes.
\cite{pi13} recently demonstrated that a variation in the initial
Zero Age Main Sequence (ZAMS) rotational velocity
(v$^{rot}_{ZAMS}$) determines a consistent spread in the final
surface s-process enhancements and spectroscopic indexes in stars
with the same initial mass and metallicity. Rotation-induced
instabilities (in particular the Goldreich-Schubert-Fricke
instability and meridional circulations) modify the mass extension
of both the $^{13}$C and the $^{14}$N pockets and their relative
overlap. This is shown in Figure \ref{fig7}, where we report the
\ct and the \nq mass fractions in the upper layers of the
He-intershell after the 4$^{th}$ TDU episode of a 2 M\odo~model
with Z=10$^{-2}$ ([Fe/H]=-0.15) and v$^{rot}_{ZAMS}$=30 km/s. We
plot chemical profiles at the end of the formation of the \ct
pocket (dotted lines) and at the beginning of the neutrons release
by the \ctan reaction (solid lines). The abundance of $^{89}$Y is
also plotted to testify the starting of neutron capture processes.
With respect to a non-rotating model the average neutron-to-seed
ratio decreases and the production of s-process elements is lower
(see also Figure 3 in \citealt{pi13}). This is a consequence of
the higher abundance of $^{14}$N, a very strong neutron poison, in
the $^{13}$C pocket. This also implies that light-s elements are
less depleted than the heavier ones. It is worth mentioning that
the inclusion of rotation does not substantially affect the
efficiency of TDU, as testified by the almost unaltered
surface [C/Fe] (see \citealt{pi13} or the FRUITY database).\\
\begin{figure*}[tpb]
\centering
\includegraphics[width=\textwidth]{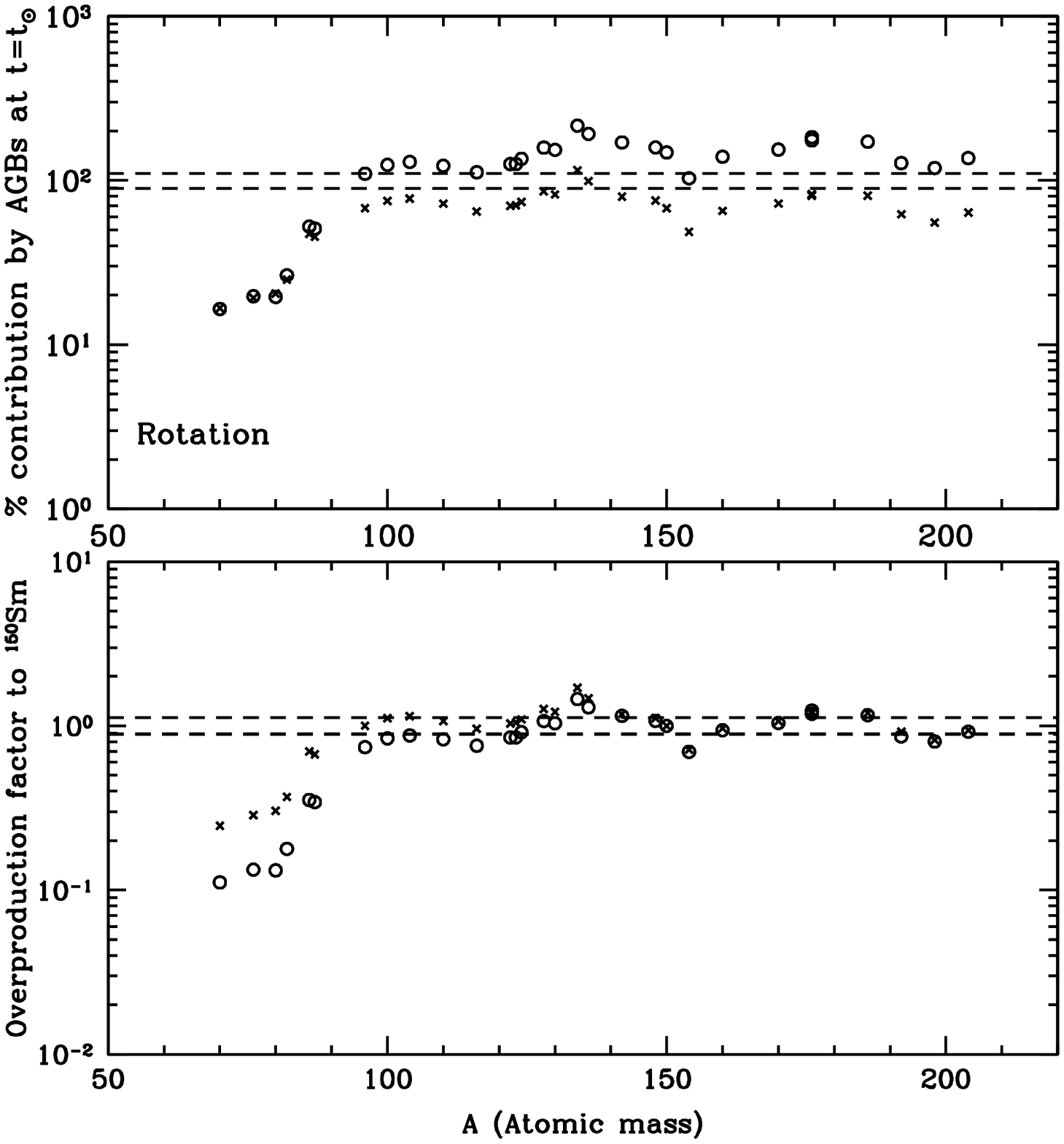}
\caption{As in Figure \ref{fig6}, but including a GCE model with
rotating AGB models (crosses).} \label{fig8}
\end{figure*}
Rotational velocities of Main Sequence stars of spectral classes A
and F span on a quite large range and they can be as high as 300
km/s. On the other hand, asteroseismology measurements seem to
indicate that the cores of Red Giant stars rotate quite slowly
(see e.g. \citealt{mosser}). This discrepancy is normally
attributed to a particularly efficient transfer of angular
momentum from the inner zones to the convective envelope or to
magnetic braking. In our models we do not account for such an
effect, but to compensate it we use low ZAMS rotation velocities.
Thus, we assume v$^{rot}_{ZAMS}$=10 km/s for models with M$\le$2.0
M\odo~and a slightly larger value for models with
2.0$<$M/M\odo~$\le$4.0 (v$^{rot}_{ZAMS}$=30 km/s). Due to the
marginal contribution to the bulk of the s-process from IMSs (see
\S \ref{refe}), we do not apply rotating corrective factors to the
yields of more massive AGBs (M$>4$ M$_\odot$). In Figure
\ref{fig8} we compare our {\it Reference} case with a GCE
calculation based on stellar models including the effects of
rotation (hereinafter {\it Rotation} case). The corresponding data
are reported in Table \ref{tab2}. As expected, we find a general
decrease of the absolute s-process abundances, the depletion
factors increasing for larger atomic masses. With the exception of
$^{138}$Ba, s-only nuclei show absolute sub-solar percentages. The
lightest s-only isotopes (up to $^{86}$Sr) are less depleted than
the heavier ones. In fact the reduced neutron exposure (due to the
partial overlap between the $^{13}$C and the $^{14}$N pockets)
leads to the synthesis of isotopes closer to the iron seeds
($^{56}$Fe), to the detriment of the heavier ones. This is even
more evident when looking to the relative overproduction factors
(lower panel of Figure \ref{fig8}). On a relative scale, light
s-only isotopes gain more than a factor 2 with respect to the {\it
Reference} case, while those in the atomic mass range 96$\le A
\le$124 are now within (or even above) the tolerance region.\\
Obviously, a different choice of the initial rotational velocities
would lead to a different s-only distribution in both the absolute
and relative scales. Thus, in principle a better fit could be
found. However, due to the other uncertainty sources affecting
stellar models, in particular those related to the treatment of
rotation in 1D evolutionary codes (see \citealt{pi13} and
references therein), we prefer to highlight general effects
related to a physical input (as rotation) more than to provide a
detailed specific recipe to obtain the desired fit.

\subsection{Convection}\label{convezione}

In our models, according to the prescriptions of the Mixing Length
Theory (MLT; \citealt{cox68}), convective velocities are
proportional to the difference between the radiative and the
adiabatic temperature gradients. Thus, in presence of a smooth
adiabatic temperature gradient profile, the convective velocity is
0 at radiative/convective interfaces. However, when the H-rich
envelope penetrates in the He-rich region, which is characterized
by a lower opacity, a non-zero convective velocity is found at the
inner border of the convective envelope\footnote{We remind that
the radiative gradient is proportional to the opacity.}. This is
the standard picture of a TDU episode. As a consequence of this
abrupt change in the opacity, the radiative/convective interface
becomes unstable \citep[see][for details]{stra06}. However, the
steep pressure gradient should limit the penetration of such an
instability and, thus, the average convective velocity should
rapidly drop to 0. We mimic this behavior by assuming that
convective velocities follow an exponential decay law below the
convective envelope. This has two major consequences: the TDU
episode is deeper and, later, a $^{13}$C pocket develops (see
\citealt{cri09} for a detailed discussion and for a comparison
with techniques used by other groups to handle the formation of
the $^{13}$C pocket).
\\
In our FRUITY models, the penetration of protons is inhibited
below 2 $H_P$ from the formal Schwarzschild Boundary (hereinafter
SB).
\begin{figure*}[tpb]
\centering
\includegraphics[width=\textwidth]{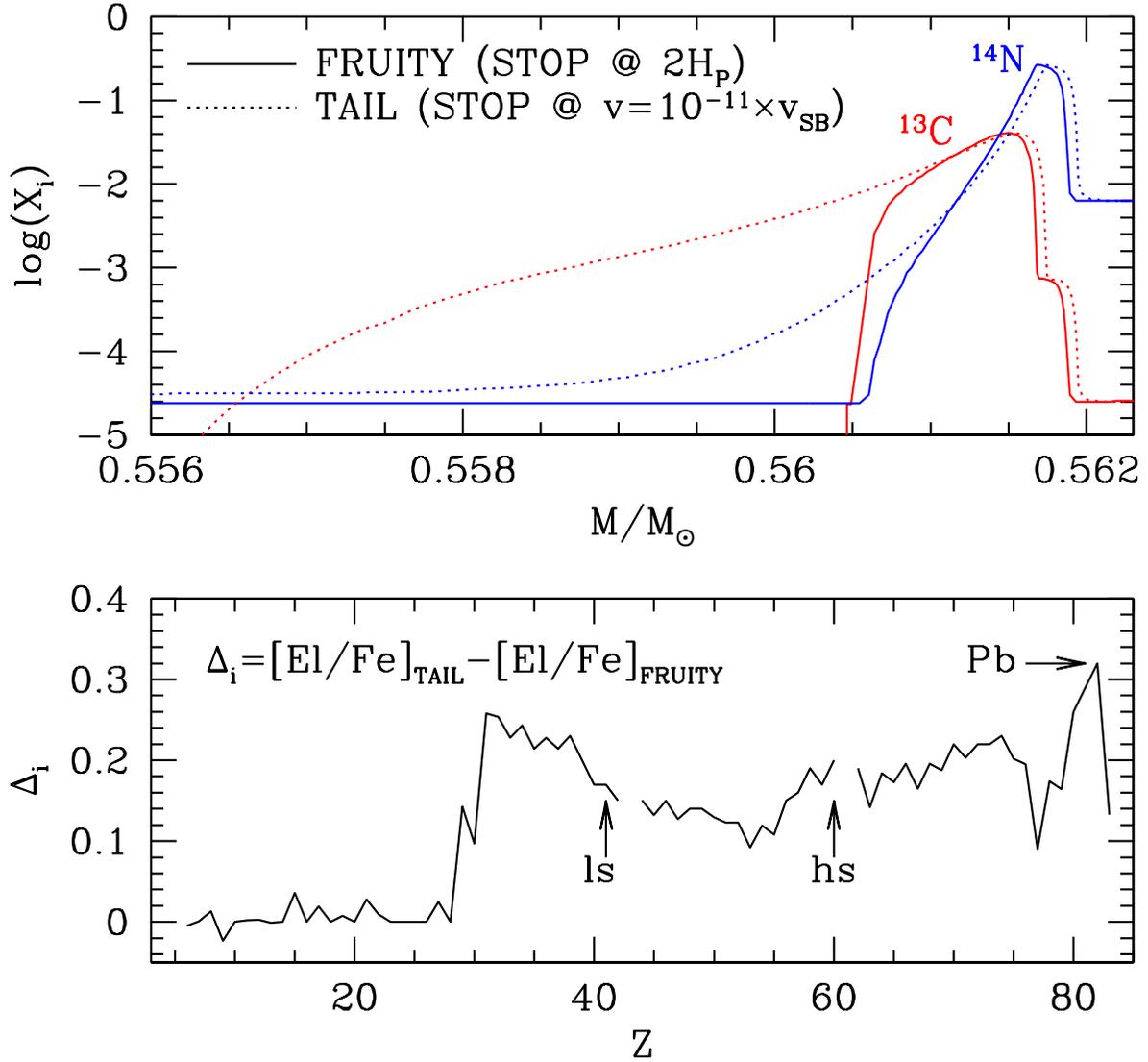}
\caption{Upper panel: $^{13}$C and $^{14}$N profiles after the
3$^{rd}$ TDU of a 2 M\odo, Z=10$^{-2}$ ([Fe/H]=-0.15) models with
different prescriptions for the radiative/convective interface
treatment. Lower panel: elemental surface differences between the
two cases shown in the upper panel. See the on-line edition for a
colored version of this figure.} \label{fig9}
\end{figure*}
\begin{figure*}[tpb]
\centering
\includegraphics[width=\textwidth]{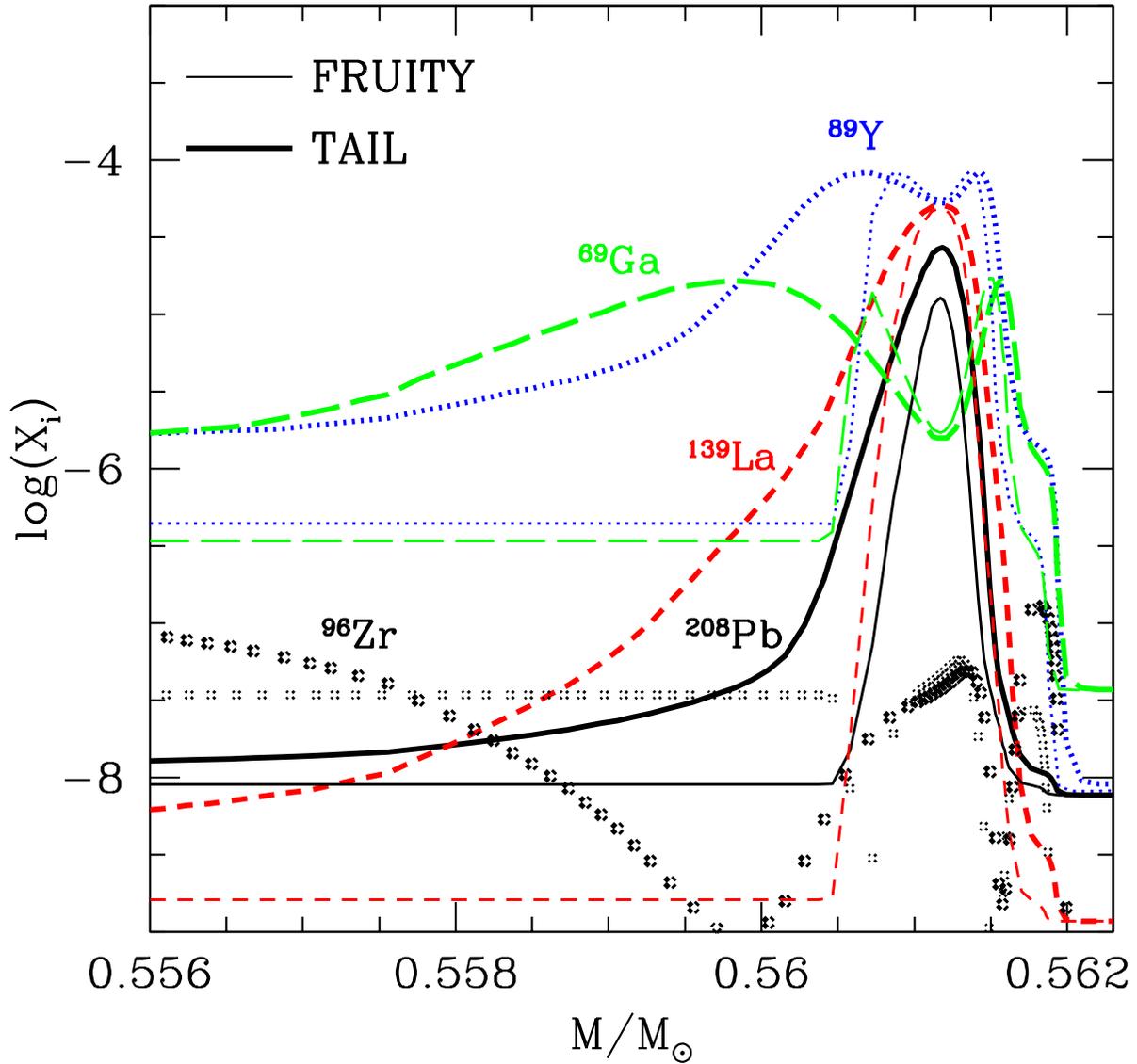}
\caption{Selected key isotope profiles in the  $^{13}$C pocket
layers after the 3$^{rd}$ TDU of a 2 M\odo~and Z=10$^{-2}$
([Fe/H]=-0.15) model. Thick and thin curves refer to the {\it
Tail} and FRUITY cases, respectively. See the on-line edition for
a colored version of this figure.} \label{fig10}
\end{figure*}
\begin{figure*}[tpb]
\centering
\includegraphics[width=\textwidth]{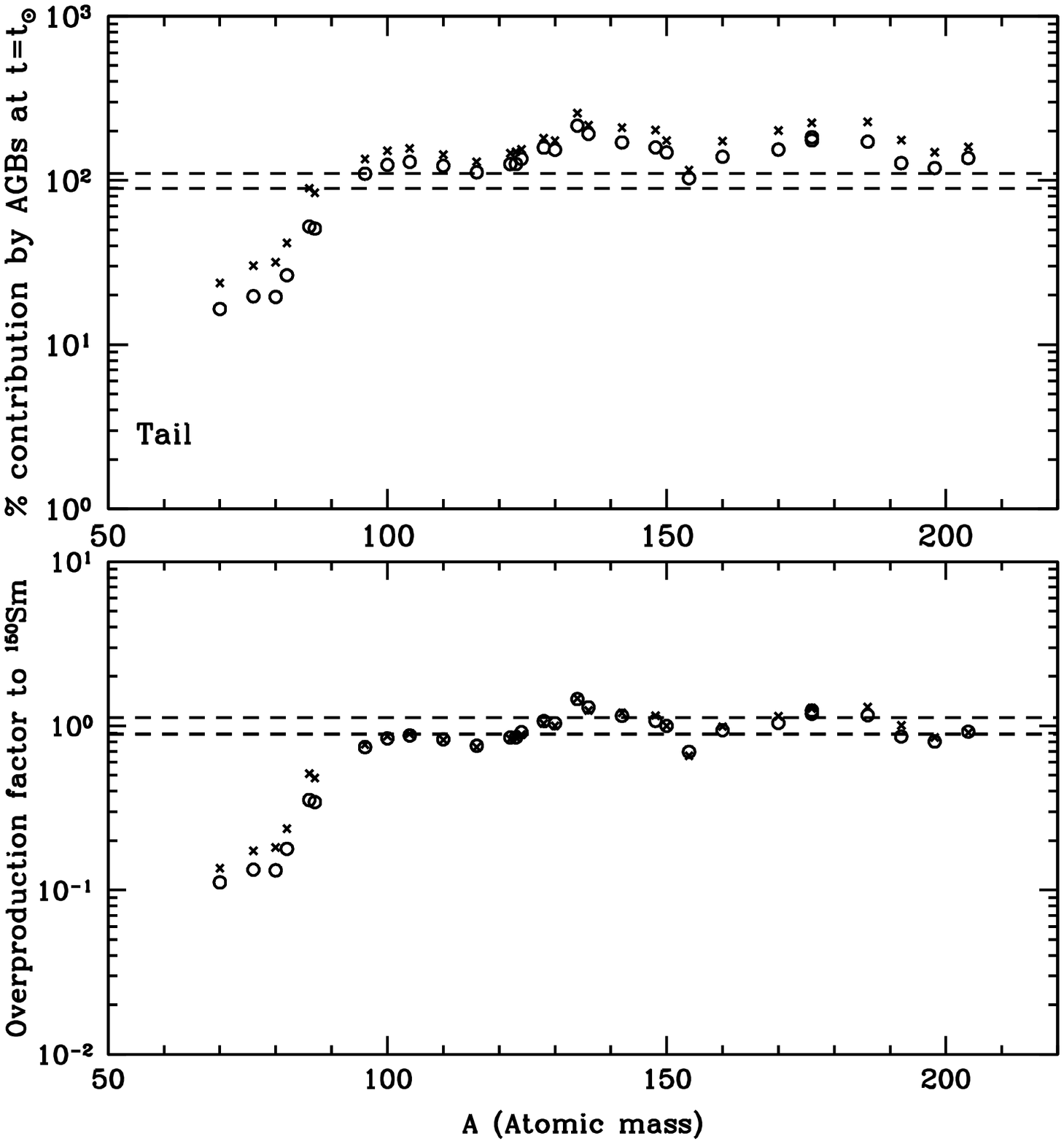}
\caption{As in Figure \ref{fig6}, but including a GCE model based
on AGB models handling in a different way the radiative/convective
interface at the base of the convective envelope (crosses). See
text for details.} \label{fig11}
\end{figure*}
In order to explore the sensitivity of stellar yield and, hence,
of GCE calculations on such an assumption on mixing efficiency, we
computed the same FRUITY non-rotating 2 M\odo~Z=10$^{-2}$
([Fe/H]=-0.15) model, but allowing the partial mixing below the SB
down to the layer where the convective velocity is 10$^{-11}$
times the value at the SB\footnote{This roughly corresponds to
(2.2$-$2.4) $H_P$.} (hereinafter {\it Tail} case). In the upper
panel of Figure \ref{fig9} we plot the \ct and \nq abundances
after the 3$^{rd}$ TDU episode of the FRUITY model (solid curves)
and the {\it Tail} model (dotted curves). As it can be easily
derived, the integrated \nq mixed below the SB is almost the same
for the two cases (7.6$\times 10^{-5}$ M\odo~vs 7.8$\times
10^{-5}$ M\odo~for the FRUITY and {\it Tail} models,
respectively), while the integrated \ct is 50\% larger in the {\it
Tail} model with respect to the FRUITY one (2.3$\times 10^{-5}$
M\odo~vs 3.4$\times 10^{-5}$ M\odo). This means that the effective
\ct \citep[i.e. the \ct that effectively contributes to the
s-process;][]{cri11} is nearly twice in the {\it Tail} model (from
9.2$\times 10^{-6}$ M\odo~to 1.8$\times 10^{-5}$ M\odo). As a
consequence, the overall s-process production increases, as
testified by the curve in the lower panel of Figure \ref{fig9},
where we plot the differences in the final surface enhancements
between the {\it Tail} and the FRUITY models. Corresponding data
are reported in Table \ref{tab2}. As expected, light elements
(Z$<28$) are not affected by the changes in the \ct tail profile,
while the three s-process peaks show larger surface enrichments
(about +30\% for ls and hs and +60\% for lead). This additional
contribution comes from the portion of the extended \ct pocket
characterized by $5\times 10^{-3}<{\rm X}(^{13}{\rm C})<1\times
10^{-2}$ (see Figure \ref{fig9}). In our previous models, such a
contribution is suppressed since it lies in correspondence to the
drop of the \ct profile. Interestingly, elements normally
associated to the weak component (Ge-Ga) result strongly enhanced
with respect to the FRUITY model (up to 60\%). This is due to the
contribution from the inner tail of the \ct pocket, where neutron
densities are lower and, thus, less massive isotopes are
synthesized (see, e.g., the $^{69}$Ga profile in Figure
\ref{fig10}). In the {\it Tail} model, neutron rich isotopes (as
$^{96}$Zr), normally bypassed by the s-process main path, are not
enhanced, but even mildly depleted with respect to the FRUITY
model. This is a consequence of the larger mass extension of the
\ct pocket of the {\it Tail} case. During the \ct radiative
burning, in fact, neutron-rich isotopes are destroyed more than
produced (see, e.g., the $^{96}$Zr profile in
Figure \ref{fig10}).\\
In the upper panel of Figure \ref{fig11} we show GCE absolute
percentage s-only abundances obtained by using AGB models with the
new prescription for the lower boundary of the \ct pocket ({\it
Tail} case). For s-only isotopes with $A\ge 96$ we find an
increase of s-process absolute percentages (on average +30\%),
with similar enhancements for light and heavy s-only isotopes (as
testified by the similar relative distributions reported in the
lower panel). Lighter s-only isotopes (A$\le 87$) result more
enhanced with respect to the heavier ones due to the contribution
from the inner tail of the \ct pocket. In summary, we find that
larger \ct-pockets do not strongly modify the shape of the s-only
distribution, but sizeably affect their absolute values (see also
Figure 4 in \citealt{bista}). We also find that the {\it Tail}
case is able to nearly reproduce the entire Galactic production of
$^{86}$Sr and $^{87}$Sr, in agreement with previous findings
\citep{trippa}. However, when compared to other s-only nuclei, Sr
s-only isotopes are underproduced and, therefore, in our GCE model
a certain contribution from the weak-s process is still
needed.\\
Note that a rigid shift (in both directions) could also be
obtained by assuming a different $\alpha$ parameter of the MLT.
Such a parameter is calibrated by reproducing the solar properties
with a Standard Solar Model (see \citealt{Piersanti2007} for
details). However, there is no specific reason to adopt the same
$\alpha$ for all the stellar evolutionary phases (see the
discussion in \citealt{stra14}). \cite{cri09} already showed that
a reduction of the MLT parameter in AGB stars leads to a decrease
of the s-process yields. Since this variation does not depend on
the metallicity, we expect a corresponding rigid shift in the
output of a GCE.

\subsection{Mass-loss}\label{reimers}

\begin{figure*}[tpb]
\centering
\includegraphics[width=\textwidth]{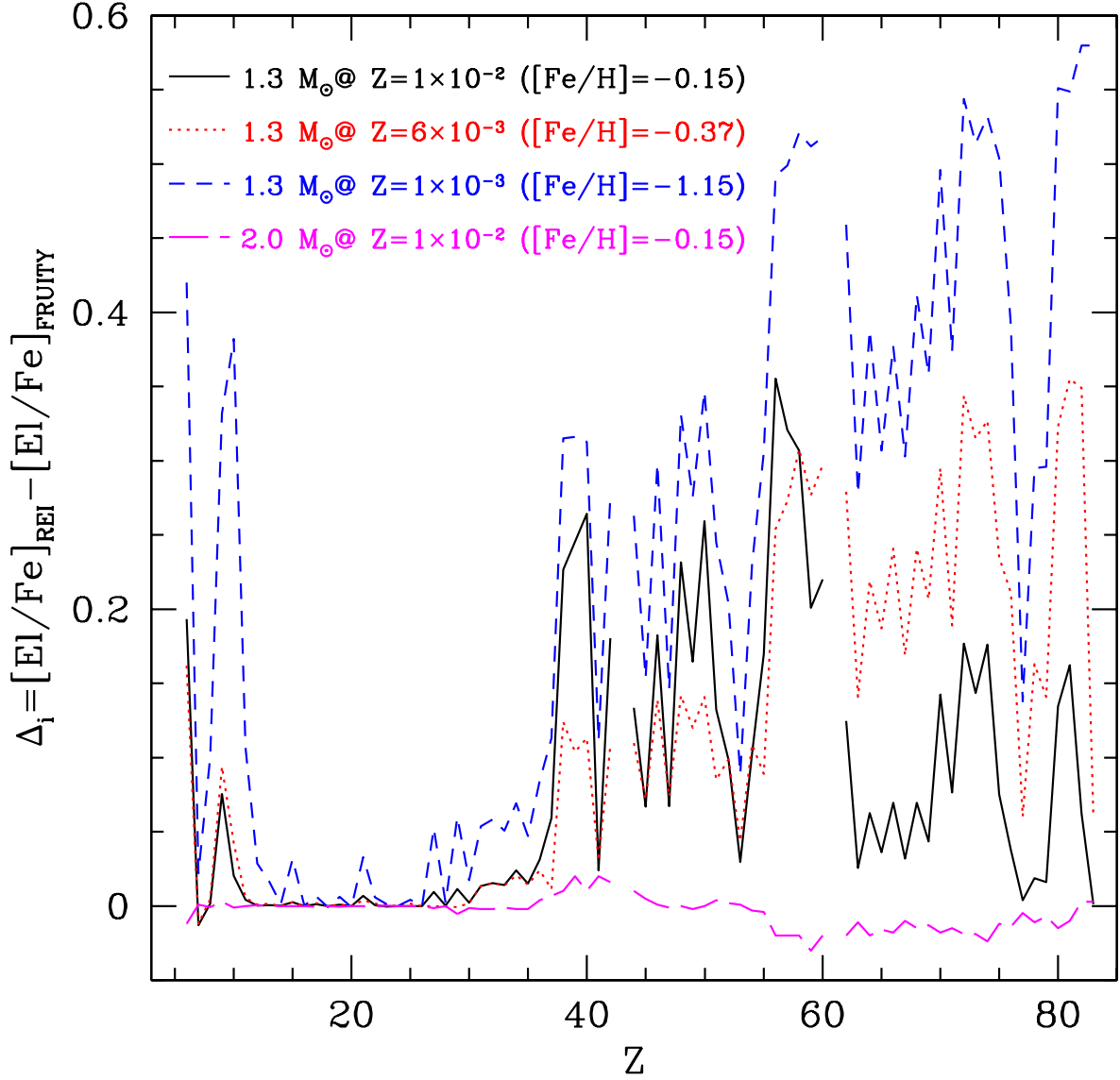}
\caption{Differences in the surface chemical distributions of 1.3
M\odo~stars at various metallicities computed with a reduced RGB
mass loss rate (REI: $\eta_R=0.2$) with respect to the
corresponding FRUITY models (FRUITY: $\eta_R=0.4$). A 2.0
M\odo~star is also reported by comparison. See the on-line edition
for a colored version of this figure.} \label{fig12}
\end{figure*}

The poor theoretical knowledge of the stellar mass loss history
represents one of the main uncertainties in the computation of AGB
stellar models. Low and intermediate mass stars lose the majority
of their mass during the Red Giant Branch (RGB) and
the AGB phases.\\
In 1D stellar evolutionary codes, the mass-loss rate during the
RGB phase is commonly parameterized according to the formulation
proposed by \cite{reimers1975}:
\begin{equation} \label{reime}
\dot M = 4\times 10^{-13} \frac{L}{g R}
\end{equation}
where $\dot M$ is in units of M\odo/yr and other quantities are in
solar units. The uncertainty affecting this formula was originally
quoted by Reimers to be at least a factor 2 either way. Later,
\citet{fpr76} introduced a normalization constant in order to
reproduce the Horizontal Branch (HB) morphology of Globular
Clusters ($\eta_R$=0.4\footnote{FRUITY models adopt this value.}).
Depending on the mass lost during the RGB phase (and thus on the
value of $\eta_R$), stars attain the AGB phase with different
envelope masses. Thus, in principle, the RGB mass-loss could have
an effect on the subsequent AGB nucleosynthesis. Those effects are
expected to be important for low mass stars (M$<$1.5 M\odo)
because they spend more time on the RGB phase with respect to
larger masses. Moreover, their envelopes are thinner with respect
to more massive stars and, therefore, even a small amount of
material lost (e.g. 0.1 M\odo) can produce sizeable effects on the
occurrence of TDU in the subsequent AGB phase (see e.g.
\citealt{stra03}). To properly determine the effects of RGB
mass-loss rate on AGB nucleosynthesis (and thus on the solar
s-only distribution), we calculate a set of M=1.3 M\odo~ models at
different metallicities with $\eta_R =0.2$. In Figure \ref{fig12}
we report the variations of the surface abundances ($\Delta_i$)
with respect to the corresponding FRUITY cases. We find that
$\Delta_i$ are larger at low metallicities (in particular for the
heaviest s-only isotopes). This is due to the fact that at large
metallicities this mass experiences a few TDU episodes even using
a milder RGB mass loss rate. Thus, the final surface s-process
enhancement is, in any case, low. By comparison, we also report
data relative to a M= 2 M$_\odot$, Z=10$^{-2}$ ([Fe/H]=-0.15)
model. The low variations found in this case confirm that for
massive enough AGB stars a reduced mass-loss rate during the RGB
phase has practically no effect.
\begin{figure*}[tpb]
\centering
\includegraphics[width=\textwidth]{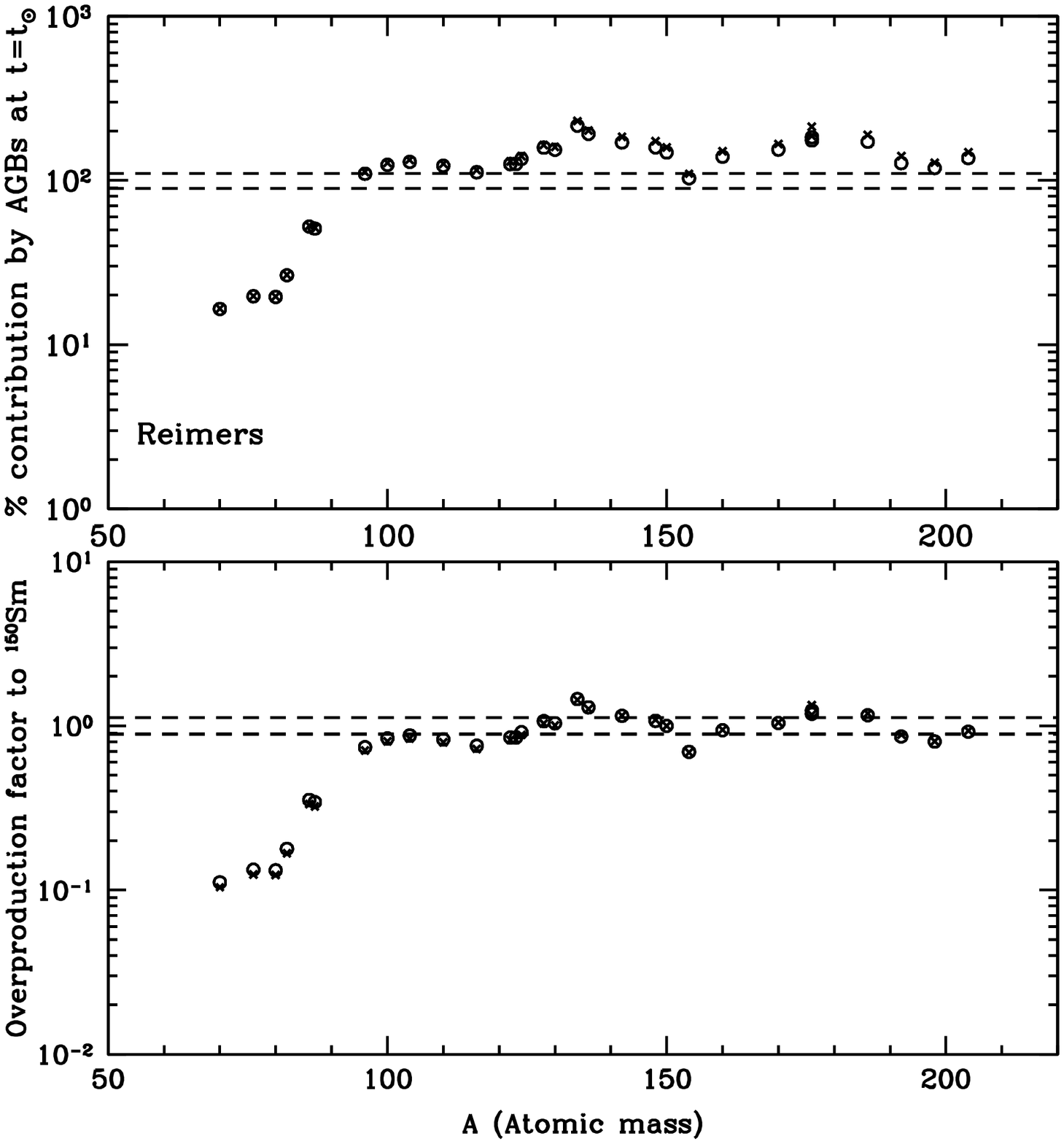}
\caption{As in Figure \ref{fig6}, but including a GCE model with a
reduced RGB mass-loss rate in low mass stars (crosses).}
\label{fig13}
\end{figure*}
In Figure \ref{fig13} we report a GCE model computed with $\eta_R
=0.2$ in stars with M$\le 1.3 $M\odo~(hereinafter {\it Reimers}
case). We find minor variations in the s-only distribution (see
also Table \ref{tab2}), with slightly larger enhancements for the
heaviest s-only isotopes ($A>128$). Our results reinforce the
evidence that the major contributors to the Solar System s-process
inventory are AGB stars in the mass range (1.5$-$3.0) M\odo, as
already inferred in Section \ref{modagb}. Their nucleosynthesis is
strongly affected by the rate at which they lose mass during the
AGB. A viable method to estimate AGB mass loss is based on the
observed correlation with the pulsation period \citep{vw93}. Since
the evolution of the pulsation period depends on the variations of
radius, luminosity and mass, this relation provides a simple
method to estimate the evolution of the mass loss rate from basic
stellar parameters. In our models, the AGB mass loss is determined
according to a procedure similar to the one adopted by
\citet{vw93}, but revising the mass loss-period and the
period-luminosity relations, taking in to account more recent
infrared observations of solar metallicity AGB stars (see
\citealt{stra06} and references therein). It has been demonstrated
that AGB mass loss rates are mildly dependent on the metallicity
\citep{groene07,laga08} and, thus, we applied the same period-mass
loss relation for all AGB models present in the FRUITY database.
\begin{figure*}[tpb]
\centering
\includegraphics[width=\textwidth]{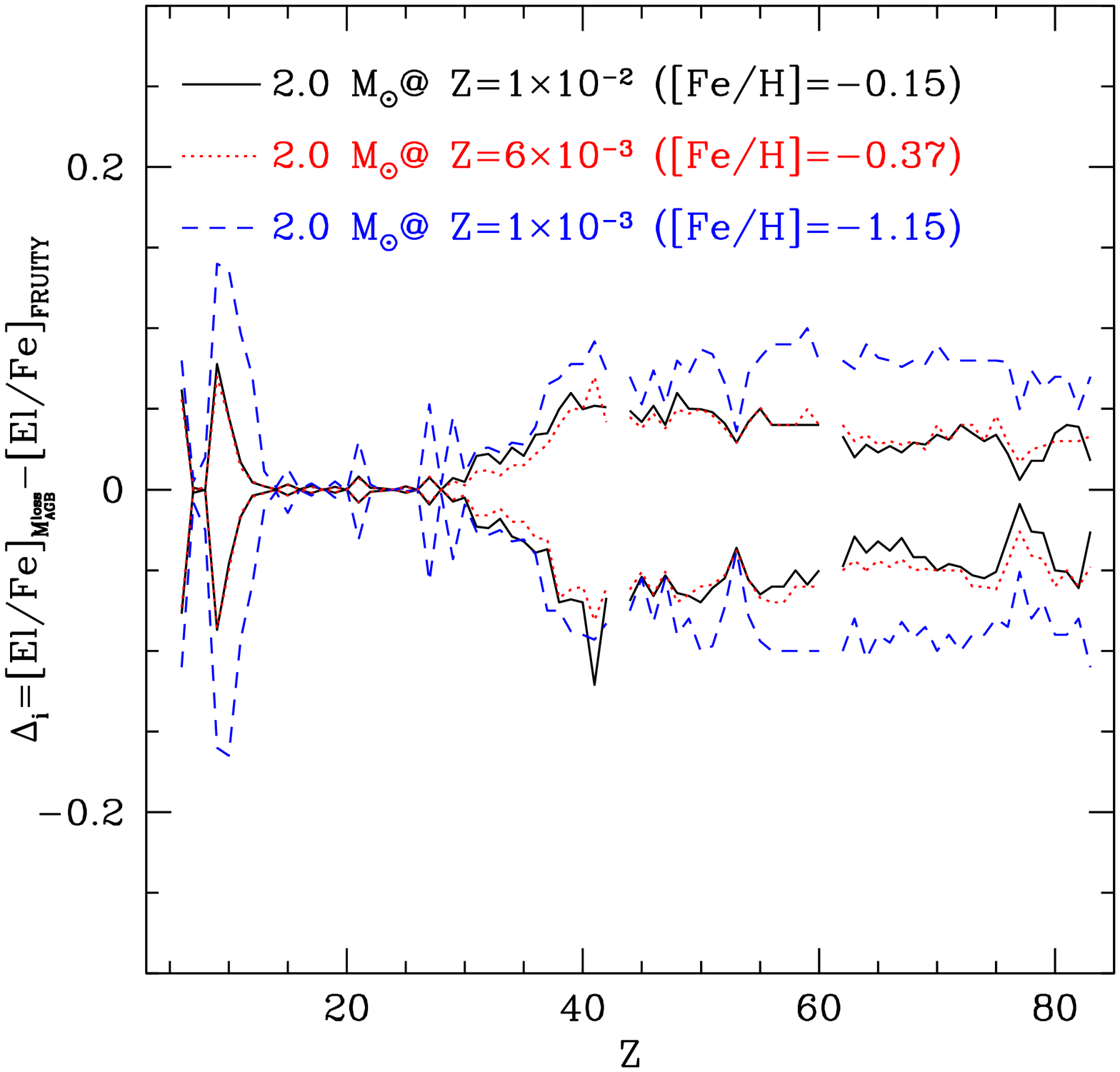}
\caption{Differences with respect to FRUITY models in the final
surface chemical distributions of 2.0 M\odo~stars at various
metallicities computed with an increased or decreased AGB mass
loss rate. Our standard AGB mass-loss rate has been described in
\citealt{stra06}. Negative differences are obtained with an
increased mass-loss rate (specularly, positive differences are
found for models with a reduced AGB mass-loss rate). As expected,
elements whose AGB production is negligible show null differences
(i.e. they have the same final surface abundances of the
corresponding FRUITY models). See the on-line edition for a
colored version of this figure.} \label{fig14}
\end{figure*}
\begin{figure*}[tpb]
\centering
\includegraphics[width=\textwidth]{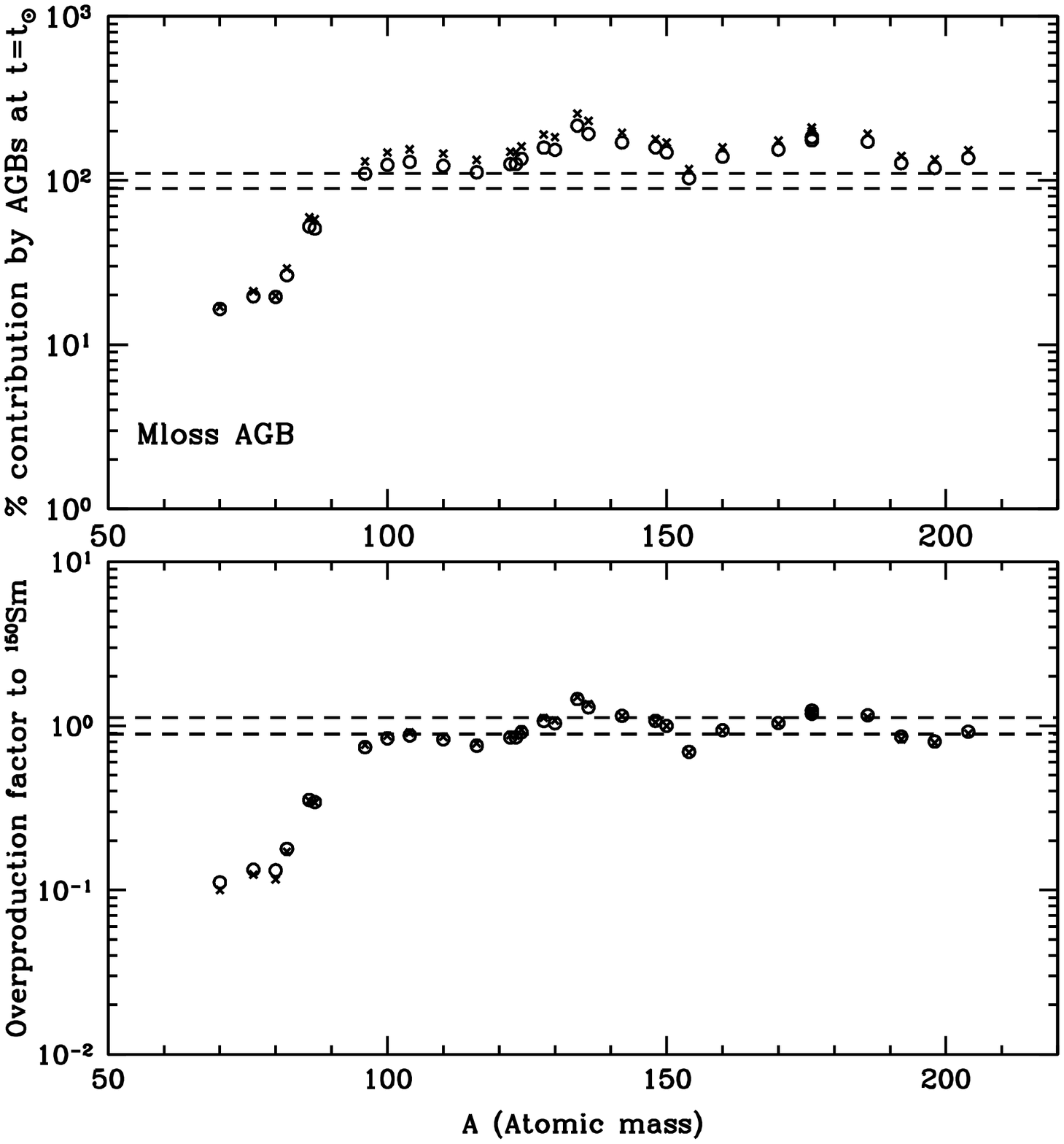}
\caption{As in Figure \ref{fig6}, but including a GCE model
obtained by assuming a lower AGB mass loss rate (crosses). See
text for details.} \label{fig15}
\end{figure*}
Notwithstanding, it is worth to note that, when a fixed period is
defined, observational data show a quite large scatter. In a
period-mass loss plot, a theoretical curve constructed reducing by
a factor 2 the mass-loss rate at a fixed period still lays within
the observed spread (see Figure 8.10 of \citealt{cristallophd}).
This still holds for a mass-loss rate increased by a factor 2. In
order to quantify the effects on the s-only distribution induced
by a variation of the AGB mass-loss rate, we compute some AGB
models with a milder and stronger period-mass loss relations. In
Figure \ref{fig14} we show the results on the final surface
distributions of 2 M\odo~stellar models at various metallicities.
In the plot, $\Delta_i$ represents the difference between models
computed with the standard $\dot M$-period relation \citep{stra06}
and the modified ones. Obviously, positive differences are
obtained with a milder mass loss rate, while negative differences
with the stronger one. Heavy elements surface variations are below
0.1 dex (25 \%) for the whole s-process distribution, being
slightly larger at low metallicity. Thus, we expect that a
modified $\dot M$-Period relation in the AGB phase produce an
almost rigid shift (upward or downward, depending on the adopted
mass loss law) of the s-process isotopes. In order to verify this
statement, we compute a GCE model with a milder $\dot M$-Period
relation during the AGB phase (hereinafter {\it Mloss AGB} case).
Results are shown in Figure \ref{fig15}; corresponding data are
reported in Table \ref{tab2}. As expected, for s-only isotopes
with A$\ge$96 there is an almost rigid upper shift of solar
percentages ($\sim$25\%). In summary, a rigid shift (upward or
downward) of the s-process isotopic inventory can be obtained by
adopting a different prescription for the AGB mass loss rate
within the intrinsic observed scatter in the $\dot M$-period
relation.

\subsection{Efficiency of nuclear processes}\label{nuclear}

In previous Sections we demonstrated that different prescriptions
on physical processes can lead to appreciable variations of the
s-only inventory. In this Section we concentrate on strong and
weak nuclear processes. We refer to \cite{cri11} for a list of the
adopted reaction rates in FRUITY models. Here, we focus on the
uncertainties affecting the rates of:
\begin{itemize}
{\item nuclear processes determining the abundances of s-only
isotopes close to s-process branchings;}{\item neutron
sources in AGB stars, i.e. the \ctan and the \nean
reactions;}{\item the major neutron poison in AGB stars, i.e. the
$^{14}$N(n,p)$^{14}$C reaction.}
\end{itemize}
By means of the first test we can quantify local variations of
s-only isotopes, while the others allow us to determine if nuclear
processes are able to shape the whole s-only distribution.\\
\begin{figure*}[tpb]
\centering
\includegraphics[width=\textwidth]{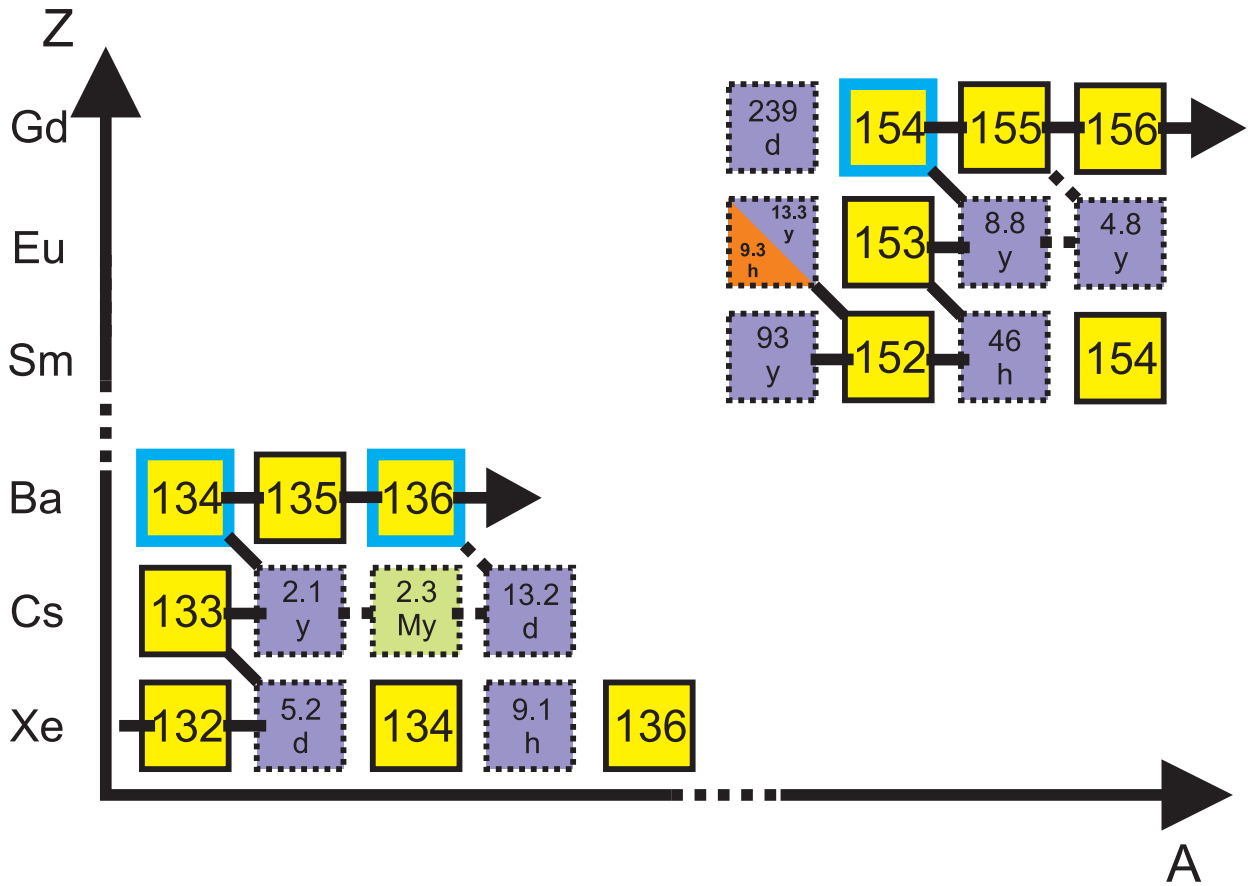}
\caption{s-process main path in the region of $^{134}$Cs and
$^{154}$Eu branching points.} \label{fig16}
\end{figure*}

\subsubsection{s-process branchings}\label{branch}

We focus on the branchings at $^{134}$Cs and at $^{154}$Eu, which
determine the surface abundances of $^{134}$Ba and $^{154}$Gd
(overproduced and underproduced with respect to $^{150}$Sm in our
GCE models, respectively). In Figure \ref{fig16} we report the
main s-process path in the regions of the nuclide chart
corresponding to the two s-process branching points. The unstable
isobars have $\beta$ decay timescales of the order of years (2.1
yr and 8.8 yr in laboratory conditions for $^{134}$Cs and
$^{154}$Eu, respectively). Thus, their decays are faster than
corresponding neutron capture during the radiative \ct burning,
but long enough to allow the opening of s-process branchings
during the convective \nean burning. Direct measurements of the
$^{134}$Cs(n,$\gamma$)$^{135}$Cs reaction is prohibitive
\citep{patronis2004}, while for the neutron capture on $^{154}$Eu
only a dated activation measurement is available \citep{ande81}.
We explore the effects of varying their neutron cross sections by
adopting the uncertainties recently provided by
\cite{rauscher2012} ($\pm$10\% and $\pm$50\% for $^{134}$Ba and
$^{154}$Gd, respectively). The $\beta$ decays rates are taken from
\cite{takayokoi1987}, while the corresponding
uncertainties (a factor 3) are taken from \cite{goriely99}. \\
\begin{figure*}[tpb]
\centering
\includegraphics[width=\textwidth]{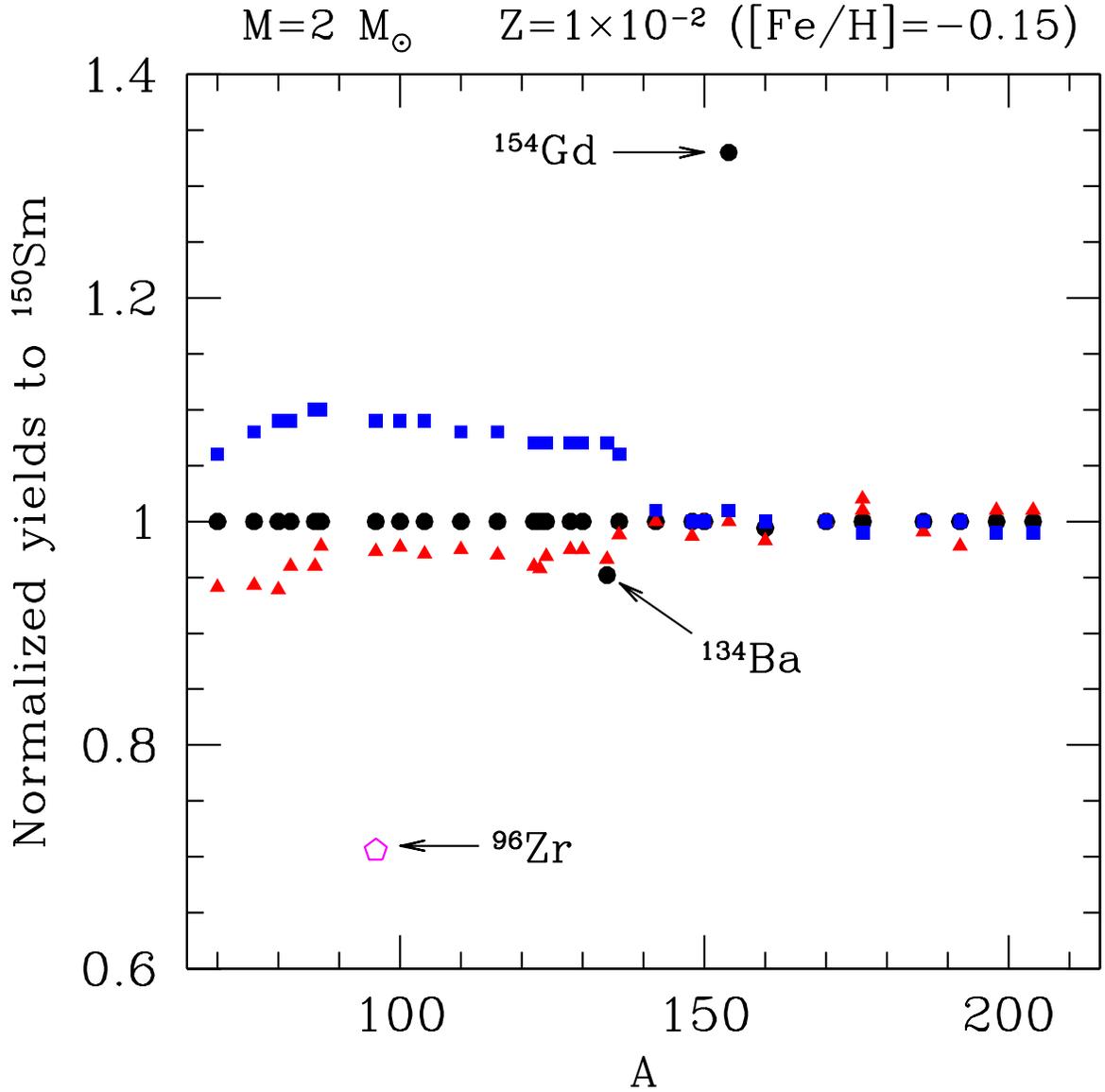}
\caption{Differences in the yields of a 2 M\odo~model with
Z=10$^{-2}$ ([Fe/H]=-0.15) obtained by varying strong and weak
reaction rates (dots: variations of nuclear processes efficiencies
in correspondence of $^{134}$Cs and $^{154}$Eu branching;
triangles: variations of neutron source cross sections;
squares: variations of neutron poisons cross sections).
Differences are normalized to variations in $^{150}$Sm yields. See
text for details.} \label{fig17}
\end{figure*}
\begin{figure*}[tpb]
\centering
\includegraphics[width=\textwidth]{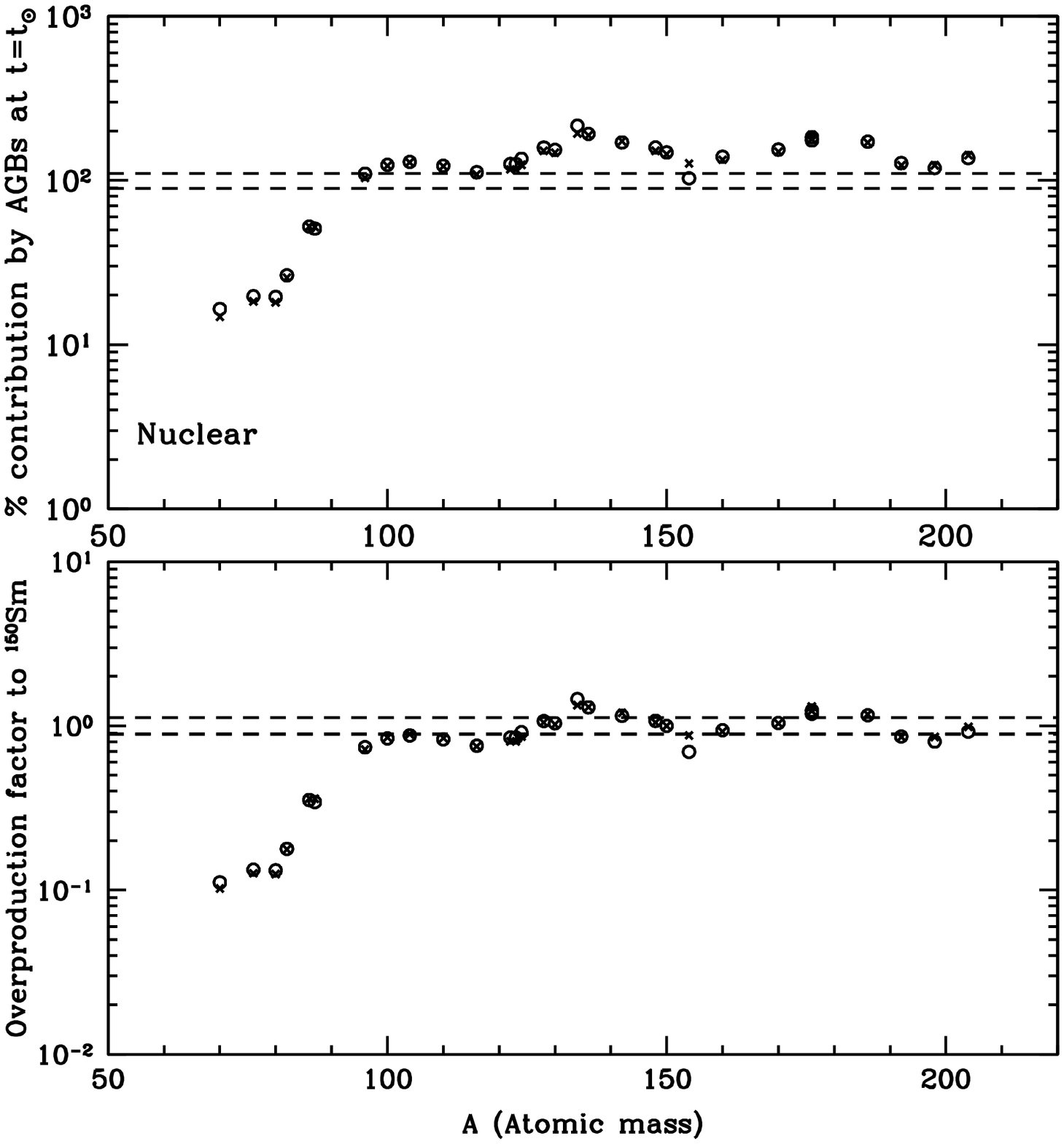}
\caption{As in Figure \ref{fig6}, but including a GCE model with
different prescriptions on selected strong and weak nuclear
processes (crosses). See text for details.} \label{fig18}
\end{figure*}
In Figure \ref{fig17} we report the differences (filled dark
circles) in the yields of a 2 M\odo~Z=10$^{-2}$ ([Fe/H]=-0.15)
model with respect to a FRUITY model by modifying neutron cross
sections and $\beta$ decays in the following way:
\begin{itemize}
\item{upper limit of the $^{134}$Cs neutron capture cross
section}; \item{lower limit for the $\beta^-$ decay rate of
$^{134}$Cs into $^{134}$Ba;\footnote{This corresponds to adopt the
upper limit of the $^{134}$Cs lifetime, which is defined as the
inverse of $\lambda$ (and thus of the rate).}} \item{lower limit
of the $^{154}$Eu neutron capture cross section;} \item{upper
limit for the $\beta^-$ decay rate of $^{154}$Eu into $^{154}$Gd
.}
\end{itemize}
These choices aim at minimizing the $^{134}$Ba production and at
maximizing the $^{154}$Gd production. The plotted quantities are
normalized to variations in $^{150}$Sm yields (thus, unity means
no variation with respect to $^{150}$Sm). We find a 5\% reduction
of $^{134}$Ba yield and a 30\% increase of $^{154}$Gd yield. Note
that these numbers refer to 2 M\odo~models: for more massive AGB
stars (e.g. 3.0-4.0 M\odo) these effects are larger.\\

\subsubsection{s-process neutron sources}\label{source}

Then, we verify if the solar s-only distribution is modified when
adopting recently published rates for the two major neutron
sources in AGB stars, i.e. the \ctan and the \nean reactions.
While the first reaction releases neutrons in radiative conditions
during interpulse periods, the latter burns in a convective
environment during TPs. For the \ctan reaction we used the value
proposed by \cite{laco2013}, while for the \nean the value
suggested by \cite{longland2012} is adopted. With respect to our
reference rates (\citealt{drotleff} and \citealt{kapp94},
respectively), both of them are about 20\% higher at the
temperatures of interest. The combined effect induced by the new
\ctan and \nean reactions is an overall slight increase of the
whole s-only distribution. This derives from the fact that with an
higher \ctan reaction rate the \ct fully burns in radiative
conditions, while when using the reference rate some of the \ct in
the first pockets can be engulfed in the convective shells
generated by TPs \citep{cri09}. When \ct is engulfed and burns
convectively, only neutron-rich isotopes as $^{60}$Fe and
$^{96}$Zr are synthesized. This is confirmed by the strongly
reduced $^{96}$Zr abundance (open pentagon) we obtain in this
model\footnote{A similar decrease is also found for $^{60}$Fe.},
despite the increased \nean reaction rate. On a relative scale, we
notice a marginal decrease of the lighter s-only isotopes with
respect to the heavy ones (triangles in Figure \ref{fig17}). The
increase of the \nean reaction does not produce sizeable effects
on the $^{134}$Ba production (see also \citealt{liu13}), which is
at the same level of s-only isotopes with 96$\le$A$\le$130. Such a
result further confirms that this reaction is only marginally
activated in low
mass AGB stars. \\

\subsubsection{s-process poisons}\label{poison}

Major neutron poisons in AGB stars are the $^{14}$N(n,p)$^{14}$C
and the $^{26}$Al(n,p)$^{26}$Mg reactions, working in \ct pockets
and during TPs, respectively. In consideration of the weak
activation of the \nean neutron source, we concentrate on the
first reaction only. Our reference rate is taken from \cite{koe}.
In Figure \ref{fig17} we report the variations in s-only isotopes
yields (squares) by considering an increased rate of 10\%. We find
a general decrease of heavy s-only isotopes, which translates in a
general overproduction of light s-only isotopes with respect to
$^{150}$Sm (+8\% on average). This is due to the fact that an
increased poison effect reduces the s-process efficiency and,
thus, its capability to by-pass the bottleneck at Z=50. Specular
results are expected when considering the lower limit of the
$^{14}$N(n,p)$^{14}$C reaction. Note that this effect is less
relevant for higher masses (in which the \nean is more efficiently
activated) and it practically vanishes at low metallicities (where
the Z=50 bottleneck is more easily by-passed due to the larger
neutrons-to-seeds ratio).

\subsubsection{Effects on a GCE model}\label{nucgcemod}

In Figure \ref{fig18} we report the results of a GCE model in
which we take into account for all of the afore-described modified
rates (hereinafter {\it Nuclear} case). Corresponding data are
reported in Table \ref{tab2}. Variations in the cross sections of
neutron sources and of the major neutron poison in AGB stars do
not lead to significative changes in the global s-only isotopes
distribution. However, this is not the case for the solar
abundances of s-only isotopes close to s-process branchings. In
fact, we find that the solar $^{134}$Ba and $^{154}$Gd percentages
decrease and increase by more than 20\%, respectively. However, we
remark that the uncertainties in the $\beta$ decay rates (which
mainly determine the differences showed in Figures \ref{fig17} and
\ref{fig18}) are rough estimates and, thus, larger isotopic
variations cannot be {\it a priori} excluded. A further
theoretical nuclear analysis on this topic would be highly
desirable.

\section{Galactic Chemical Evolution Models Uncertainties}\label{chiesto}

It is important to remind that also GCE models strongly depend on
the adopted inputs, such as the SFR, the IMF or the type Ia
supernovae evolutionary scenario (see Section \ref{refe}). In
fact, each of these quantities influences the amount of metals
locked or released by stars at different epochs. The relevance of
their impact would depend on how much their variations (within the
current uncertainties) would affect the derived age-metallicity
relation. For instance, a faster increase of the ISM metallicity
would imply a lower contribution from metal-poor stars, because
there would be less time to form them. Thus, the contribution from
metal-rich AGB stars to the Solar System s-process distribution
would be larger and, consequently, the production of s-only
isotopes with $96\le A \le 124$ would result increased
(\citealt{maiorca}; see also \citealt{trippa}). In fact, the
higher the iron seeds number the lower the atomic mass of the
synthesized s-process nuclei. We plan to systematically study the
impact of different choices of the GCE input parameters on the
s-only distribution in a forthcoming paper. Hereafter we only show
the effects that a variation of the SFR has on the solar s-only
isotopic
distribution. \\
\begin{figure*}[tpb]
\centering
\includegraphics[width=\textwidth]{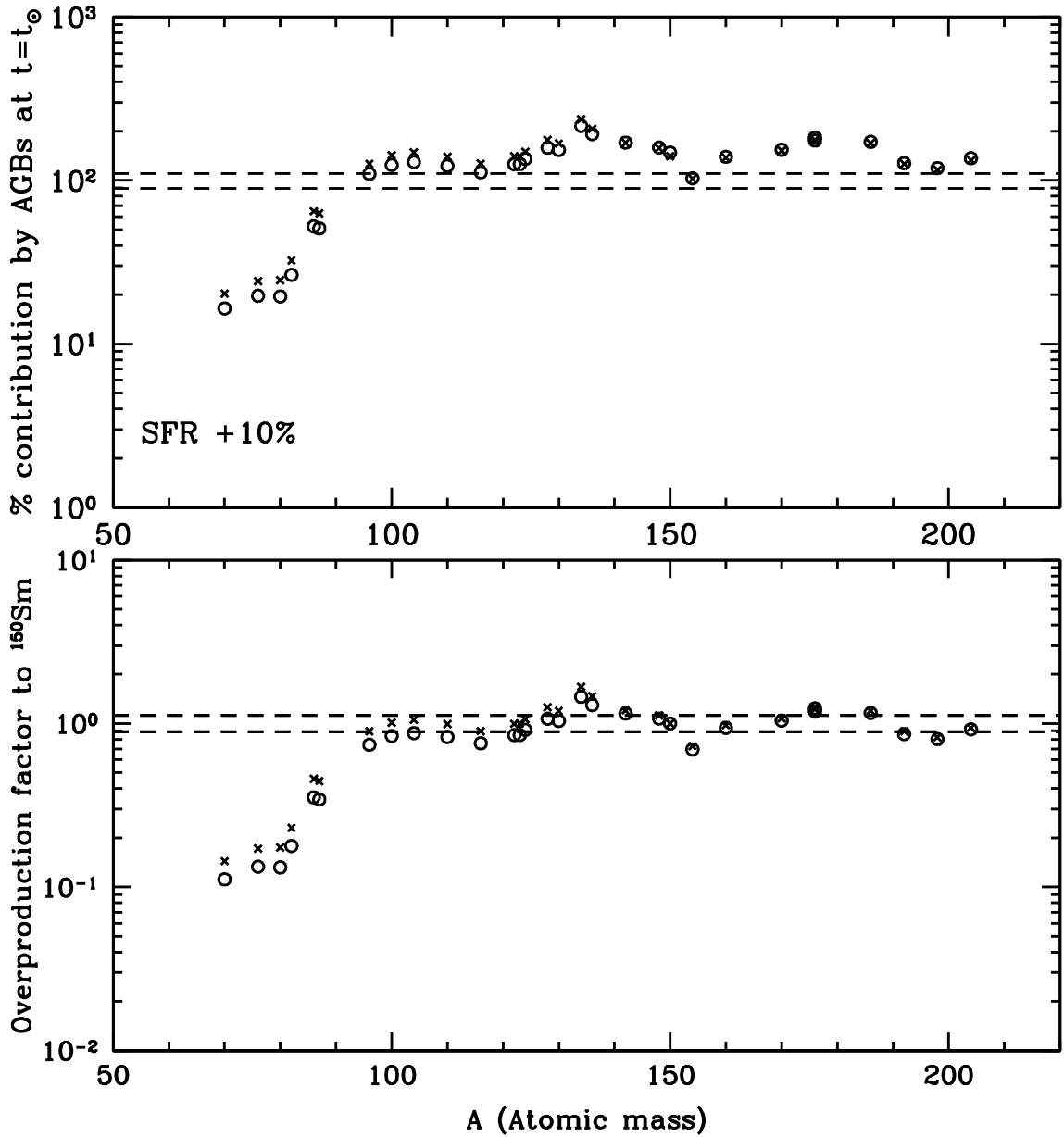}
\caption{As in Figure \ref{fig6}, but including a GCE model with
an increased SFR of +10\% (crosses). See text for details.}
\label{fig19}
\end{figure*}
Observations of various SFR indicators in galaxies reveal that
star formation occurs in different ways, depending on the galaxy
type. There is no theory to predict star formation on large scales
in a galaxy given the many physical ingredients that may affect
the SFR. In Figure \ref{figc1} we show the age-metallicity
relation obtained by assuming an increased SFR at all epochs
(+10\%; dashed curve) with respect to our {\it Reference} case.
Here an increased SFR mimics a higher [Fe/H]. Corresponding data
are reported in Table \ref{tab2}. We will refer to this case as
SFR+10. Actually, we have just changed the $\alpha$ parameter in
the Schmidt's law by $10\%$, because with this choice it is still
possible to account, within observational uncertainties, for all
the solar neighborhood observables mentioned in Section
\ref{refe}. A larger variation of the SFR would imply a new
calibration of the GCE model itself. In this case, however, it
would be difficult to disentangle the effects related to the
change in the SFR from those connected to the new parameter set
adopted to fit again observables. As shown in Figure \ref{fig19},
the variation of the SFR has an appreciable effect on the s-only
isotopes distribution. We notice a slight increase of light s-only
isotopes and a more consistent decrease of the heavy ones. As a
consequence, on a relative scale light s-only isotopes with $96\le
A\le 136$ are overproduced with respect to the heavier ones by
18\% on average.

\section{Discussion and Conclusions} \label{conclu}

In this paper we verify if our FUNS stellar yields (available on
the FRUITY database), used in a Galactic chemical evolution model,
can reproduce the distribution of s-only isotopes characterizing
the proto-solar nebula. Those nuclei are only synthesized by the
s-process and, thus, are exceptional markers of the evolution of
past Galactic AGB populations. At odd with previous studies based
on post-process calculations \citep{bista,trippa}, we use in our
analysis AGB stellar yields obtained by means of stellar
evolutionary
calculations fully coupled to an extended nuclear network.\\
In our GCE models, we find that the contribution to the Solar
System s-only distribution from low mass AGB stars (M$<$ 1.5
M\odo) as well as from intermediate mass AGB stars (M$>$ 4 M\odo)
is marginal. Thus, we confirm that the bulk of the s-process comes
from AGB stars with masses ($1.5-3.0$) M\odo. Another major result
of this study is that, within the combined uncertainties, we do
not miss any contribution to the Solar System s-only distribution
in the atomic mass range $96\le A \le 124$, as claimed by
\cite{travaglio04} and \cite{bista}. Our reference GCE model, in
fact, predicts an overall super-solar s-only distribution ($\sim
+45$\% on average). When observational and nuclear errors are
taken into account, the distribution relative to $^{150}$Sm can be
considered flat, even if a lower production is found for s-only
nuclei with $96\le A \le 124$. We investigate if current
uncertainties affecting stellar models can lead to a better fit to
the Solar System s-only isotopic distribution. The inclusion of
rotation in our stellar models implies a general suppression of
the s-process, with larger depletion factors for the heaviest
s-only isotopes. On a relative scale, this implies a larger
contribution to light s-only isotopes and, thus, a flatter s-only
distribution. Different prescriptions for convection efficiency
and for the treatment of the unstable inner border of the
convective envelope during TDU episodes produce nearly rigid
shifts of the entire s-only distribution. The same result can be
achieved by adopting a different mass-loss rate during the AGB
phase. Current nuclear uncertainties affecting strong and weak
reactions allow for important improvements in the determination of
some s-only isotopes (as $^{134}$Ba and $^{154}$Gd). The need of
revised $\beta$ decay rates with respect to those published by
\cite{takayokoi1987} is
highly compelling. \\
In the past, the nucleosynthesis of s-only isotopes has been
closely related to that of $^{208}$Pb. Although such a nucleus is
not a pure s-process isotope, a large percentage of its solar
abundance is ascribed to the s-process, the estimates varying from
about 85\% \citep{cowan99} to 98\% \citep{bista}. Our reference
model slightly overestimates its absolute solar abundance (108\%);
as a consequence, about 27\% of $^{208}$Pb is missing with respect
to $^{150}$Sm (which has an absolute percentage solar abundance of
148\%). Taking into consideration its still uncertain s-process
contribution and the observational error in the determination of
its solar abundance, we are missing about 10\% of solar $^{208}$Pb
at minimum. Note, however, that at odds with the s-only isotopes
studied in this paper, this isotopes could receive a non
negligible contribution from very low metallicity AGB stars (see
Figure \ref{fig0}), which are not taken into account in our
simplified GCE model. Thus, we can assume our $^{208}$Pb
production as a sort of lower limit. Concerning the test models
previously discussed, we find that the absolute abundance of
$^{208}$Pb roughly scales as the $^{150}$Sm one. On a relative
scale, minor variations ($<$5\%) are found in the majority of
tests, apart from the {\it Rotation} case (-9\%) and the {\it
Tail} case (+22\%). The latter could be a good candidate to
compensate the relative $^{208}$Pb underproduction found in the
{\it Reference} case. \\
It is important to remark that, in addition to the uncertainties
of AGB stellar models here discussed, other uncertainties may
affect the predicted s-only distribution. As it is well known, AGB
stars at various metallicity contribute differently to the three
s-process peaks. Thus, if the contribution from stars at large Z
is favored \citep{trippa}, a flatter relative s-only distribution
may be found. Thus, the hypothesis on the existence of a LEPP
process also relies on the uncertainties currently affecting
Galactic chemical evolution models. We verified that an increase
of the Star Formation Rate at all epochs leads to a faster
increase of the ISM metallicity and, thus, to a larger
contribution from metal-rich stars. As a consequence, we obtain a
larger production of light s-only isotopes with respect to the
heavy ones and, consequently, a flatter distribution.\\
In conclusion, our full stellar evolutionary models coupled to a
GCE model for the solar neighborhood does not necessarily require
the need for a LEPP mechanism to be able to increase the Solar
System s-only abundances in the range $96\le A \le 124$. However,
owing to the uncertanties still affecting both stellar and
Galactic chemical evolution model, we cannot {\it a priori}
definitely rule out the existence of additional contributions to
the Solar System s-only isotopes distribution. Note that the
models presented in this paper cannot certify (or rule-out) the
existence of a metal-poor primary LEPP, invoked to explain the
abundances of a large group of light elements in low metallicity
halo star which might be enriched by an r-process.

\acknowledgments

We thank the referee for valuable comments and suggestions, that
improved the quality of this paper. This work was supported by
Italian Grants RBFR08549F-002 (FIRB 2008 program), PRIN-MIUR 2012
"Nucleosynthesis in AGB stars: an integrated approach" project
(20128PCN59) and from Spanish grants AYA2008-04211-C02-02 and
AYA-2011-22460.

\bibliographystyle{aa}
\bibliography{cristallo}

\begin{deluxetable}{rlcccccccc}
\tablecolumns{8} \tablewidth{0pc} \tablecaption{Absolute
percentage isotopic abundances with respect to the solar
distribution (see text for details). Solar percentage errors
(taken from \citealt{lo09}) are also reported (column 2). }
\tablehead{\colhead{\footnotesize Isot.} & \colhead{\footnotesize
{$\delta_\odot$(\%)}} & \colhead{\footnotesize {\it Reference}} &
\colhead{\footnotesize {\it No IMS}} & \colhead{\footnotesize {\it
Rotation}} & \colhead{\footnotesize {\it Tail}} &
\colhead{\footnotesize {\it Reimers}} & \colhead{\footnotesize
{\it Mloss AGB}} & \colhead{\footnotesize {\it Nuclear}} &
\colhead{\footnotesize {\it SFR+10}}
\\} \startdata
 $^{70}$Ge  &16  &   17 &   13 & 17  & 24  &  17  &  17  &  15  & 20 \\
 $^{76}$Se  &7   &   20 &   16 & 19  & 30  &  20  &  21  &  18  & 24 \\
 $^{80}$Kr  &20  &   20 &   16 & 21  & 32  &  20  &  20  &  18  & 25 \\
 $^{82}$Kr  &20  &   26 &   22 & 25  & 42  &  27  &  29  &  26  & 32 \\
 $^{86}$Sr  &7   &   52 &   49 & 47  & 90  &  53  &  59  &  52  & 65 \\
 $^{87}$Sr  &7   &   51 &   46 & 45  & 84  &  52  &  58  &  52  & 63 \\
 $^{96}$Mo  &16  &   110&   104& 68  & 135 &  113 &  131 &  104 & 126\\
 $^{100}$Ru &6   &   124&   118& 75  & 151 &  128 &  148 &  123 & 143\\
 $^{104}$Pd &11  &   130&   124& 78  & 157 &  134 &  155 &  129 & 148\\
 $^{110}$Cd &7   &   123&   117& 72  & 144 &  126 &  146 &  122 & 140\\
 $^{116}$Sn &16  &   112&   107& 65  & 130 &  115 &  133 &  109 & 127\\
 $^{122}$Te &7   &   126&   120& 70  & 146 &  130 &  150 &  117 & 140\\
 $^{123}$Te &7   &   126&   120& 71  & 149 &  131 &  149 &  116 & 141\\
 $^{124}$Te &7   &   135&   129& 74  & 155 &  140 &  161 &  124 & 150\\
 $^{128}$Xe &20  &   158&   152& 86  & 181 &  165 &  190 &  152 & 177\\
 $^{130}$Xe &20  &   154&   147& 82  & 175 &  160 &  184 &  148 & 168\\
 $^{134}$Ba &18  &   216&   208& 115 & 257 &  230 &  255 &  193 & 236\\
 $^{136}$Ba &18  &   192&   185& 99  & 217 &  202 &  230 &  188 & 207\\
 $^{142}$Nd &5   &   170&   163& 80  & 210 &  184 &  195 &  174 & 170\\
 $^{148}$Sm &5   &   159&   152& 75  & 202 &  173 &  179 &  152 & 158\\
 $^{150}$Sm &5   &   148&   140& 68  & 175 &  159 &  170 &  145 & 141\\
 $^{154}$Gd &14  &   103&   97 & 49  & 115 &  110 &  117 &  127 & 102\\
 $^{160}$Dy &15  &   139&   132& 65  & 173 &  151 &  159 &  134 & 138\\
 $^{170}$Yb &5   &   154&   147& 72  & 201 &  167 &  175 &  150 & 152\\
 $^{176}$Lu &5   &   183&   176& 83  & 224 &  186 &  210 &  189 & 177\\
 $^{176}$Hf &5   &   175&   174& 81  & 225 &  213 &  199 &  179 & 172\\
 $^{186}$Os &8   &   172&   164& 81  & 228 &  189 &  192 &  170 & 168\\
 $^{192}$Pt &8   &   128&   120& 62  & 176 &  140 &  141 &  124 & 126\\
 $^{198}$Hg &20  &   119&   112& 55  & 149 &  128 &  134 &  124 & 116\\
 $^{204}$Pb &7   &   137&   130& 64  & 160 &  148 &  152 &  142 & 133\\
 \hline
 \enddata \label{tab2}
\end{deluxetable}

\end{document}